\documentclass[%
 reprint,
 amsmath,amssymb,
 aps,
]{revtex4-2}

\usepackage{graphicx}
\usepackage{dcolumn}
\usepackage{bm}
\usepackage{hyperref}
\usepackage{xurl}
\usepackage[utf8]{inputenc}
\usepackage[T1]{fontenc}
\usepackage{mathptmx}
\usepackage{etoolbox}
\usepackage{physics}

\usepackage{epsfig}
\usepackage{amsmath}
\usepackage{appendix}
\usepackage{relsize}
\usepackage{amssymb}
\usepackage{float}
\usepackage[mathscr]{euscript}
\def\be{\begin{eqnarray}}
\def\ee{\end{eqnarray}}

\def\={&=&}
\def\f{\frac}

\def\be{\begin{eqnarray}}
\def\ee{\end{eqnarray}}

\def\L{\left[}
\def\R{\right]}
\def\={&=&}
\def\f{\frac}

\def\vs{\vspace{3mm}}

\begin{document}

\title{How hate spreads online and why it returns:\\
Re-entrant phases driven by collective behavior}

\author{Chen Xu}%
\affiliation{ 
School of Physical Science and Technology, Soochow University, Suzhou 215006, China 
}%

\author{Pak Ming Hui }%
\affiliation{ 
Department of Physics, The Chinese University of Hong Kong, Shatin, Hong Kong SAR China
}%

\author{Chenkai Xia}%
\affiliation{ 
Physics Department, George Washington University, Washington D.C. 20052, U.S.A.
}%

\author{Neil F. Johnson}%
\affiliation{ 
Dynamical Online Networks Laboratory, George Washington University, Washington D.C. 20052, U.S.A.
}%

\begin{abstract}
The 2025 Bondi Beach mass-shooting was perpetrated by individuals inspired by ISIS (Islamic State) propaganda that increasingly featured anti-Semitic hate content following the October 2023 start of the Israel-Palestine war. Similar stories hold for other types of hate attacks, e.g. against Muslims on May 18, 2026. There is an urgent need to get ahead of future threats by understanding how and when a newly created piece of hate content will spread system-wide online. We present a two-species coalescence-fragmentation model with Susceptible-Infected-Recovered dynamics that incorporates the following published empirical features: (1) New pieces of hate content tend to be generated and promoted by a subset of in-built communities on less regulated platforms. (2) These `hate' communities create links (hyperlinks) with each other and with non-hate communities across all platforms to form dynamically evolving clusters (i.e.\ coalescence) across which new hate content can then spread. (3) These clusters can get broken up by moderator shutdowns (i.e.\ fragmentation). We present numerical solutions and derive two levels of approximate mean-field theory: Effective Medium Theory (EMT) and Beyond Effective Medium Theory (BEMT). Both numerical and analytic solutions reveal that system-wide spreading is governed by re-entrant threshold phases: as the fraction of hate communities  varies, the system can transition from spreading to no-spreading and back to spreading. The derived analytic formulae give explicit insight into how these phase boundaries might be manipulated to prevent system-wide spreading. 
More broadly, the re-entrant phase behavior warns that policies which steadily reduce the number of hate communities can initially succeed but then backfire if pushed further, suggesting that blanket requirements for platforms to simply do `more' are over-simplistic.  
\end{abstract}

\maketitle

\section{Introduction}

Australian authorities have confirmed that the December 2025 Bondi Beach attack was inspired by ISIS  (Islamic State) ideology \cite{homeland,Jewish,CTV}.
After ISIS lost its territorial caliphate and the Gaza war started in October 2023, online ISIS propaganda surged and increasingly focused on anti-Semitic framing and calls for attacks on Jews. Online communities on less regulated platforms like VKontakte, Telegram and Gab, continue to create and circulate new hate content that encourages low-tech attacks against synagogues and Jewish gatherings \cite{homeland,Jewish,CTV,Multi2021,usNature2019,usScience2016,Lupu,npjAks,npjMin,city,Boogs,PRL2023,Brain,Tina}. 
In response, the Australian government has pledged to introduce tougher laws on hate speech offenses and the display of extremist symbols. Other governments are also looking closely at the growing threat within their own countries caused by this new nexus of anti-Semitic sentiment  combined with ISIS' continued ability to incite violent attacks anywhere in the world. Similar stories follow irrespective of the target of the hate, e.g. Muslims. 

However, without a concrete quantitative model for how  new pieces of hate content manage to spread, policymakers have no reliable basis for deciding  what to recommend \cite{Tina,brown,Gill,Cald,threats,examplecases,30,shapiro1,surge,UK,metaverse1}
 -- other than simply restricting online use and issuing blanket requirements for social media platforms to do `more'. But what should this concrete quantitative model for hate-content spreading actually be?

There are already many excellent works in Physics that describe general viral spreading on model networks. As an illustration of this burgeoning literature, we refer to Refs.  \cite{Watts,Newman,Barabasi,Menczer,Lamboitte,T1,C1,S1,H1,L1,K1,J1,M1,J2,I1,B1,B2,Holme,Stanley,Vespignani,HH,Gavrilets,Redner,usPRE2010,5,6,7,8,Palla07,bak2022,chenkai,2025book,PRL2024,Om} but there are many more that are equally interesting and which deserve inclusion in any full review. Though such a full review lies beyond the scope of the present paper, we note that many of the existing published works go beyond static network setups to consider different ways in which links might form and break while some viral content is being passed around \cite{Watts,Newman}. Some  incorporate different types of networks, including multilayer networks with different node and link types \cite{Newman,B1,B2}; and several include different types of viral process beyond Susceptible-Infected-Recovered (SIR) which nonetheless remains the standard \cite{Watts,Newman,Barabasi,Menczer,Lamboitte,T1,C1,S1,H1,L1,K1,J1,M1,J2,I1,B1,B2,Holme,Stanley,Vespignani,HH,Gavrilets,Redner,usPRE2010,5,6,7,8,Palla07,bak2022,chenkai,2025book,PRL2024,Om}.
However, there is to our knowledge no model that incorporates the specific empirical features of where pieces of pro-ISIS and anti-Semitic hate content originate and how they then manage to spread online \cite{Multi2021,usNature2019,usScience2016,Lupu,npjAks,npjMin,city,Boogs,PRL2023}. The present paper attempts to fill this gap.

This paper presents a rigorous mathematical analysis that reveals the conditions under which a piece of newly created hate content from within a given online community, will then spread systemwide. To achieve this, we develop a model (which is presented in detail in Sec. II)  that incorporates specific empirical features from prior published studies of the online multi-platform universe. References \cite{usScience2016,PRL2023} provide empirical evidence that the underlying scaffolding of support for ISIS and other extremist anti-Semitic entities evolves empirically in time through the coalescence of in-built communities (nodes) into clusters. Any new piece of hate
content can then spread through the roadways provided by this cluster scaffolding. In addition, there is empirical evidence that
 these clusters occasionally fragment due to moderator shutdowns \cite{PRL2023}. 
 
 Hence there is an ongoing dynamical interplay -- due to coalescence and fragmentation -- between the making and breaking of roadways through which new hate content may then pass (or not). Each of these in-built communities, which Refs. \cite{Multi2021,usNature2019,usScience2016,Lupu,npjAks,npjMin,city,Boogs,PRL2023} refer to as {\em hate communities} because they each have a history of promoting anti-Semitic hate speech as evidenced by their page's content, could for example be a Club on the platform VKontakte, a Channel on Telegram, or a Group on Gab. Figure 1 provides an empirical example of pieces of hate content spreading, with data taken from Ref. \cite{Multi2021} to which we refer for full empirical details. Each in-built community (i.e. small circle in Fig. 1(a)) has its own unique platform assigned ID; it may have hundreds of thousands of members (i.e. users); and any of the approximately 5 billion online users worldwide is either a member of a given in-built community or not. 
We stress that these in-built communities have nothing to do with community-detection in network science. A benign version of such an in-built community on Facebook, for example, would be a parenting group \cite{parents1,parents2,parents3}, sports fans of a particular team, or a pet lovers' group -- and many of us are likely members of such an in-built community on one social media platform or another. For visual simplicity, the in-built communities shown in Fig. 1 are all just hate communities.

\begin{figure}[!t]
\centering
\epsfig{figure=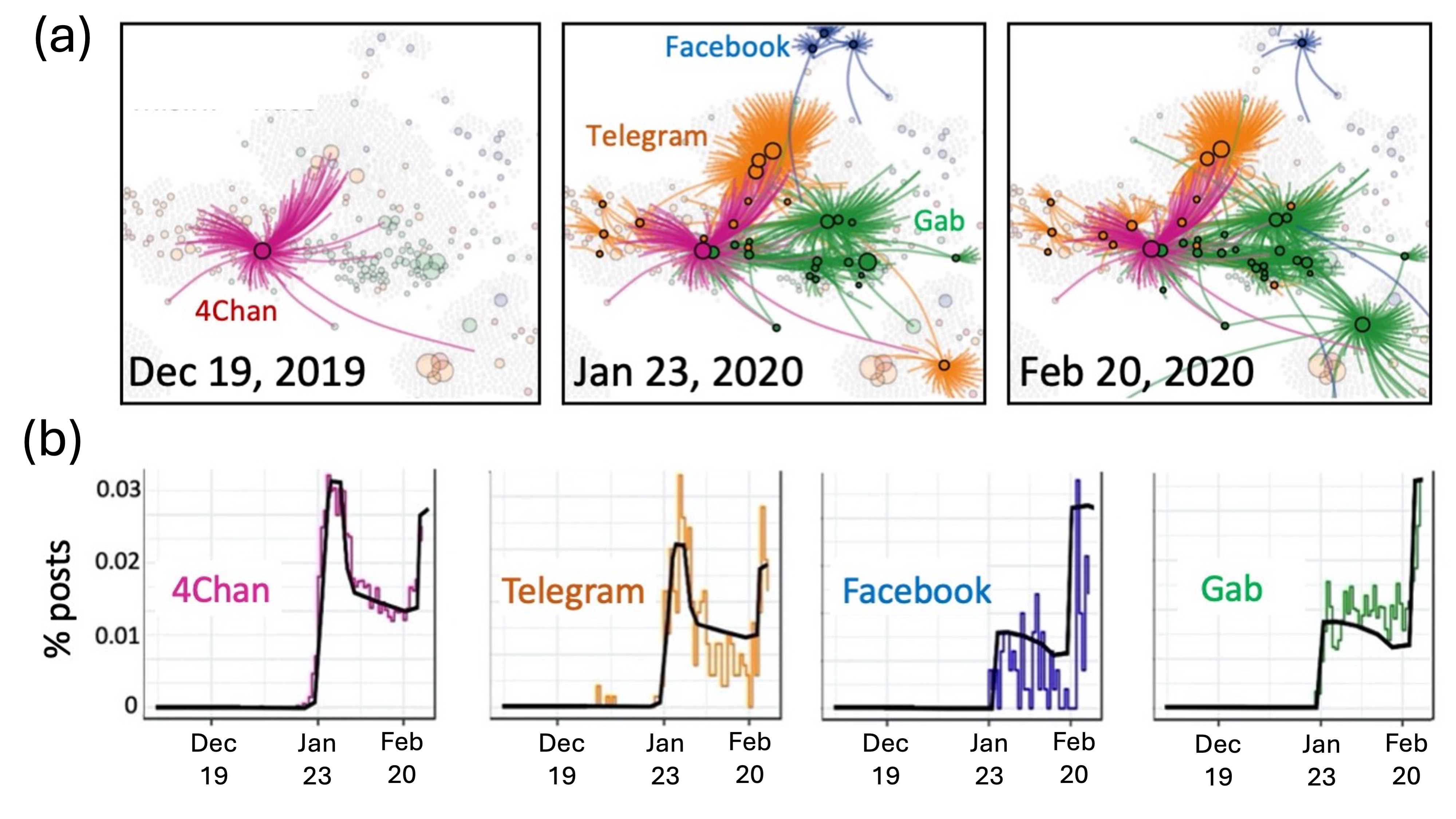,width=1.0
\linewidth} 
\caption{(a) Example of empirical spreading of anti-Semitic-focused hate speech by extremist communities (e.g. pro-ISIS communities and beyond) across different platforms. The new anti-Semitic hate content in this example is focused around COVID-19, and comes from early in the pandemic: data from Ref. \cite{Multi2021}. Each node (small circle) is a hate community on a given  platform, e.g. a Gab Group that has previously featured hate speech. Each link is a hyperlink created by a given hate community (node) to another hate community (node) \cite{Multi2021}: here we only show links that carry hate content at that instant in time. Each link's color denotes the platform of origin \cite{Multi2021}. (b) shows total amount of this hate content on a given platform from the empirical data in (a), as a percentage of all content at that instant. Curves for each platform change in a highly non-monotonic way with increasing time, suggesting re-entrant phase diagrams for system-wide spreading -- in line with the model's predictions in this paper. Though detailed empirical comparison is beyond the present paper's scope, the black curves show example outputs from the simplest version of the model in which both node species have identical parameter values (see Fig. 2 for the model's parameters).}
\label{fg:multiverse}
\end{figure}
\noindent

This paper's model obviously represents a simplified mathematical cartoon of the complexities of the real online world. But this simplicity means that we can derive approximate analytic equations for its behavior, which we then show agree with its detailed numerical simulation (see later Sections). Our model's cartoon simplicity also means that it can in principle be applied to the spreading of other types of hate speech beyond anti-Semitism, 
as supported empirically by Refs. \cite{Lupu,npjAks,npjMin} and others, which show that anti-immigration hate speech etc. also spreads in similar ways to Fig. 1. 

In terms of physics, this paper contributes to the long-standing goal of understanding transport processes on dynamical networks -- and in particular, it identifies and derives new mean-field approximations which agree well with the exact numerical simulations. 
More specifically, the paper helps advance Physics' understanding of dynamical network spreading  \cite{Watts,Newman,Barabasi,Menczer,Lamboitte,T1,C1,S1,H1,L1,K1,J1,M1,J2,I1,B1,B2,Holme,Stanley,Vespignani,HH,Gavrilets,Redner,usPRE2010,5,6,7,8,Palla07,bak2022,chenkai,2025book,PRL2024,Om} by obtaining numerical and analytic mean-field solutions for spreading on a coalescence-fragmentation network with two species of node, as explained in Sec. II. 

To position this paper precisely within the existing literature, the present study represents the natural and previously unexplored convergence of three prior published works: (1) Ref. \cite{usPRE2010} combined a single-species ($D=1$) coalescence-fragmentation model with SIR contagion dynamics on top, but only explored the results numerically without deriving analytic mean-field theories or systematically mapping out the phase diagram behavior. (2) Ref. \cite{PRL2023} established the coalescence-fragmentation framework on a rigorous empirical and theoretical footing, but did not add any contagion process such as SIR on top of the clustering dynamics, nor did it fully explore the multi-species case. (3) Ref. \cite{PRL2024} generalized the coalescence-fragmentation model to $D$ species and demonstrated that multi-species cohesion dynamics produce qualitatively new phenomena, but again did not add any contagion process on top. The present paper fills the gap left open by all three: it combines the multi-species coalescence-fragmentation dynamics (here $D=2$ species, labelled A and B) with SIR contagion, derives analytic mean-field solutions at two levels of approximation (EMT and BEMT), and systematically maps the resulting phase diagrams -- revealing re-entrant spreading phases that were not present in any of the three prior works.

The outline of this paper is as follows. Section II explains the details of the model and walks through the steps in the model's numerical simulation. The model features two species of node: hate communities (A) and non-hate communities (B). We explore three empirically motivated parameter regimes: Case~I (inter- vs.\ intra-species link formation, controlled by parameter $F$), Case~II (asymmetric species-dependent coalescence, controlled by parameter $G$), and Case~III (digital vaccination, in which B nodes have a reduced infection rate $\chi = \phi\beta < \beta$, where $0 \leq \phi < 1$). 
Section III presents the model's approximate analytical analysis and solution by means of mean-field approximations at two levels: Effective Medium Theory (EMT) and Beyond Effective Medium Theory (BEMT). Section IV presents and compares results from the model's exact numerical simulation with its approximate solutions from these mean-field approximations. The simulations reveal -- and the analytic solutions confirm -- the existence of 
re-entrant threshold  behaviors for the transition between a phase where there is no system-wide spreading and a phase where there is system-wide spreading. Section V draws these results together in a compact way to discuss policy implications for controlling system-wide spreading of hate content.

\section{Model and its Numerical Simulation}

\subsection{Empirical motivation and our approximations}
Our goal in this paper is to develop and study a minimal model of how a new piece of hate content spreads, where the minimal model's features are consistent with the known empirical features of hate content spreading from prior empirical studies; and then to develop approximate mean-field descriptions of the complex phase boundary behaviors that emerge from the model's numerical simulations. As we show in the subsequent sections, this is already quite daunting and detailed. Hence we do not carry out any additional quantitative comparison with empirical data, since this would involve addressing additional intricacies of the online world and human behavior. However, we do offer in Fig. 1(b) some reassurance that the model can reproduce similar non-monotonic spreading (i.e. infection) profiles to the real online world. Specifically, 
the black curves show example outputs from the simplest version of the model in which both species of node (i.e. in-built community) have identical parameter values. We will present detailed comparisons between these empirical infection profiles in time and those from the model in future works, but in the rest of this paper we will focus exclusively on establishing the emergent behaviors of the theoretical model alone.

\vskip0.1in

Hence our challenge here is to design a minimal model that incorporates, in some simple yet manageable way, the following known and published empirical features: 
\vskip0.1in

\noindent (1) Large amounts of anti-Semitic hate content are created within, and promoted by, a subset of in-built  communities that tend to reside on less regulated, so-called fringe platforms. Examples of such a `hate community' include a VKontakte Club, a Telegram Channel, a Gab Group. Following Refs. \cite{Multi2021,usNature2019,usScience2016,Lupu,npjAks,npjMin,city,Boogs,PRL2023,Brain}, any in-built community that has explicitly featured anti-Semitic hate content in the past is labeled as a `hate community'. We refer to these prior empirical studies for detailed discussions of how hate speech and hate crimes are defined in legal terms and how far back in the past to consider.
In terms of the in-built communities themselves, numerous studies confirm that individuals aggregate online into in-built communities around a particular common interest (e.g. a Facebook Group or Page, a VKontakte Club) in order to share advice and experiences \cite{parents1,parents2,parents3}. We all tend to do this ourselves -- and ISIS supporters and other anti-Semitic extremists are no different \cite{usNature2019,usScience2016}. Each resulting in-built community can be represented as a node in a network. 
\vskip0.1in

\noindent (2) These hate communities (nodes of type A) then create hyperlinks (links) with each other over time, and with some of the more mainstream in-built communities across all platforms (e.g. Facebook Group) to form dynamically evolving clusters (i.e.  {\it coalescence}) as seen in the empirical data in Fig. 1(a). These hence generate time-evolving roadways upon which the virus (i.e. new hate content) can then get added. Specifically, a hate community (node A)
on a given social media platform establishes a link (by sharing a URL
for a piece of content) to another community (node) at a given time. A link from
node 1 to node 2 means that members of community 1 are alerted to 2's existence,
and can visit community 2 to share hate content. Likewise members of community 2 will become aware of community 1's existence through the comments of the community 1 visitors, and hence may themselves visit community 1 to see its hate content. As a result, any new hate content can easily spread within any given cluster at any given moment since there is, by definition, a path available between any two nodes in the cluster at that moment. 

Empirically, the number of non-hate communities involved (i.e. the B nodes in our model) is only a small subset of the billions of existing online non-hate communities (nodes). This is because the vast majority of non-hate communities are so benign that they never get connected to hate communities (A nodes) either directly or indirectly, and hence they are completely irrelevant to the hate content spreading process -- so they are not included in the model. But a small subset of these non-hate communities do get involved at some stage, either through a direct link or by being in a cluster with hate communities (A nodes): this small subset are the B nodes in our model. Some B nodes may be pet lover communities, but only a small fraction of all pet lover communities are B nodes. They may conceivably eventually convert into hate communities: but we do not concern ourselves in this paper with possible longer term conversion of B nodes to A nodes. 
All this means that the number of hate nodes A and non-hate nodes B involved in the spreading process can be comparable empirically, which is why we pick similar population numbers for A and B nodes in some of the results that we show.

\vskip0.1in

\noindent (3) The resulting clusters of A and B nodes can get broken up by occasional moderator shutdowns that have been observed empirically (i.e. {\em fragmentation}). Reference \cite{PRL2023} for example, shows an explicit empirical example of such a total cluster fragmentation event. This fragmentation tends to occur infrequently because a cluster's anti-Semitic activity first needs to be noticed and reported, and then it requires moderator effort to apply sufficient pressure (e.g. threat of sanctions). There are also empirical examples of a cluster effectively self-fragmenting as a result of its nodes deciding to self-silence their links in order to avoid attention, or the links simply becoming inert -- which has the same net result as forced fragmentation in terms of hindering spreading. There could also be partial cluster breakup -- but as long as the resulting fragments are small then the spreading results will be qualitatively similar to those presented here: we refer to Ref. \cite{2025book} for a full discussion of such partial fragmentation. Our model focuses on complete fragmentation because this immediately prevents that cluster from spreading new hate  content.
\vskip0.1in

These empirical observations (1)-(3) motivate us to develop a simple model of A and B nodes aggregating into clusters because of  links being added over time between in-built communities (nodes) within and across platforms -- and sporadic fragmentation of these clusters. As well as enabling the joining together of existing clusters of any size, such coalescence also includes the simple case of one of the clusters having 
size one (i.e. isolated node) and hence the second cluster will simply appear to grow by one node.
Of course, such a coalescence-fragmentation model could be implemented in myriad ways with varying levels of real-world complexity. For example, it is quite possible in the real online world that the nodes (in-built communities) that form links to each other have some multi-vector character that makes such link formation more likely; and that the path that a particular piece of anti-Semitic content follows depends on such intrinsic characteristics of the nodes as well as prior history. It is also quite possible that the contagion process whereby a node receives new content and decides to share it more widely, is also a function of the content itself. However, we will ignore such complications in order for our model to be tractable: indeed, understanding the effect of adding such additional complications requires first understanding the system without them. Hence we will assume that links get added randomly, and that the probability that a cluster is chosen for fragmentation depends simply on its size (i.e. the number of nodes that it has) in order to capture how noticeable it is.

Though obviously a cartoon of the actual online network dynamics, empirical support for these simplifications includes an analytic solution in the limit of small fragmentation probability \cite{EPL2025,ben,scirep2013,Nature2009,Oxford,Dylan1,Dylan2,PRL2023,2025book} to yield an approximate power-law cluster distribution with power-law exponent $-5/2$, which is in agreement with the empirical findings for pro-ISIS support online \cite{usScience2016,PRL2023,EPL2025,ben,scirep2013,Nature2009,Oxford,Dylan1,Dylan2,2025book}. This is not the same as the well-known percolation threshold which happens to appear at a particular value of the link-to-node ratio as links are added: instead, it is a self-organized steady-state phenomenon in a system in which clusters evolve dynamically, but in which the largest cluster will only infrequently make it to the percolation threshold (and will only be there temporarily) because of the fragmentation process which, by picking nodes randomly, favors larger clusters. Hence whether of not the largest clusters are above the percolation threshold or not, is in practice irrelevant since they only remain that size for a small period of time which is insufficient for widespread spreading to occur across it. In short, this is not the typical static network percolation spreading situation, and the system-wide spreading in our system is not dominated by the percolation threshold. We also note that our coalescence-fragmentation mechanism connects to the classical Smoluchowski coagulation framework with a product kernel, but now with complete fragmentation \cite{Redner,Stockmayer1943,Lushnikov1978,DaviesKingWattis}.

To allow readers to explore the richness of the resulting two species (A and B node) coalescence-fragmentation cluster dynamics interactively --- before any viral spreading process is added --- we provide a free, publicly accessible online simulation dashboard at \url{https://gwdonlab.github.io/netlogo-simulator/}. The dashboard implements the full multi-species model with independently adjustable coalescence and fragmentation probabilities for each species of node. Running the simulation, the following can be seen in real time: (a)~the network of nodes and clusters evolving visually, with mixed A and B node clusters forming through coalescence and breaking apart through fragmentation; (b)~the cluster size distribution which changes in time, and a comparison to the approximate $-5/2$ power law; (c)~the size of each individual cluster tracked over time, showing the characteristic coalescence and fragmentation events; (d)~the size of the largest cluster fluctuating in a dynamical steady state; and (e)~the species composition of the giant connected component (i.e. the number of A and B nodes within it) at each timestep. A representative screenshot is shown in Appendix~B (Fig.~\ref{fg:netlogo}). 

\vskip0.2in
We now discuss the contagion process that we add on top of these coalescence-fragmentation cluster dynamics. We consider the standard choice of SIR (Susceptible-Infected-Recovered).
This is of course a huge simplification compared to the nuances of actual online content and people's decisions about it. We could in principle use the same underlying coalescence-fragmentation dynamical network model in combination with many different variants of the contagion process, such as Susceptible-Infected-Susceptible. However, each such exploration deserves its own study beyond the scope of the present paper.
We will additionally consider the possibility of digital vaccination, in which the non-hate (B) nodes have a reduced infection rate $\chi = \phi \beta$ where $0 \leq \phi < 1$, reflecting the empirical observation that some platforms are already attempting to ``inoculate'' vulnerable mainstream communities against hate content by preemptively posting counter-messages \cite{bak2022}. The parameter $\phi$ controls the vaccine efficacy: $\phi = 0$ corresponds to perfect vaccination (B nodes are immune) while $\phi = 1$ means no vaccination effect. This is explored in Case~III.

\vskip0.2in

\subsection{Details of model's numerical simulation}
We present now the step-by-step details of the numerical simulation of our coalescence-fragmentation--plus--SIR model. Section III then treats the model analytically using approximate mean-field theory. Section IV's comparison to the detailed numerical simulation confirms that the exact numerical simulations and the approximate mean-field solutions are in reasonable agreement. The fact that this agreement arises despite the coarse-graining of the mean-field equations, suggests that the specific choices implemented in  the numerical simulation are not too important, i.e.  it suggests that our findings of re-entrant phases will be fairly robust to variations in the numerical simulation details.

\begin{figure}[!t]
\centering
\epsfig{figure=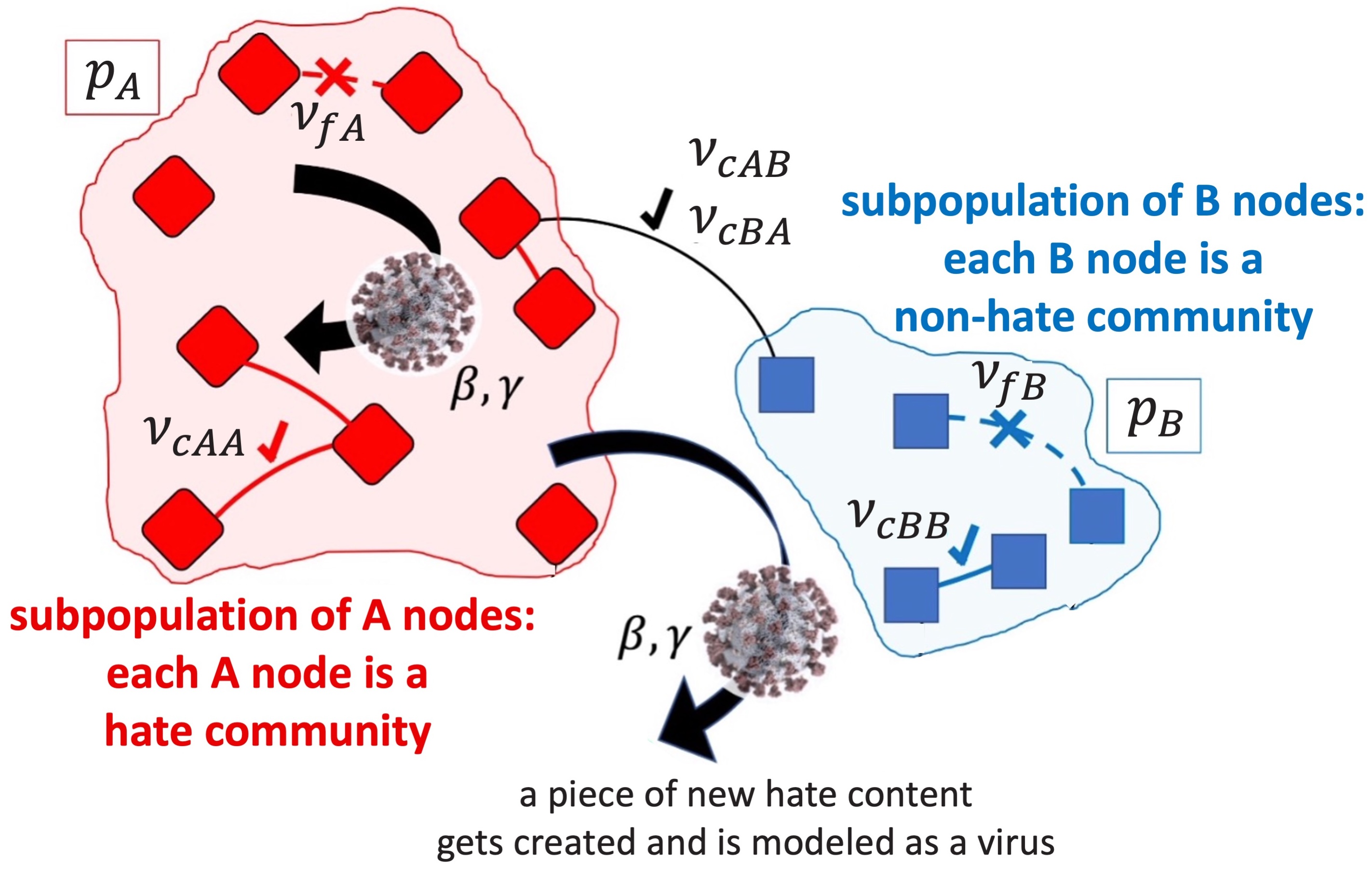,width=1.0
\linewidth} 
\caption{Our coalescence-fragmentation--plus--SIR model shown in a static schematic form. It features the coalescence-fragmentation dynamics of the two species of nodes: hate communities (A) and non-hate communities (B), which provides the time-evolving roadways through which new hate content can then spread (akin to a new virus). For a real-time visualization, see our online simulation dashboard at \url{https://gwdonlab.github.io/netlogo-simulator/} and choose the two-species option. At any given time, there are clusters of mixed sizes with mixed species compositions, i.e. A and B nodes. Using the mean-field version of this dynamical model as derived in Sec. III, we can crudely mimic the empirical results in Fig. 1(b) using a minimal set of parameter values: $\nu_{cAA}=\nu_{cBB}=\nu_{cAB}=\nu_{cBA}=0.95$, $\nu_{fA}=\nu_{fB}=0.05$, $\beta=0.05$, and $\gamma=0.01$ for panels 1-2 in Fig. 1(b)  and $\gamma=0.005$ for panels 3-4.}
\label{fg:schematic}
\end{figure}
\noindent

Our numerical simulation of the coalescence-fragmentation--plus--SIR model proceeds as follows. At any time $t$, users in any in-built community (node) 1 may create a link (hyperlink) with any other in-built community 2 (node) \cite{Multi2021,npjAks,npjMin}. This can then expose the linked community 1 to any new hate content that community 2 might have and hence this hate content can spread to community 1's members. This link also alerts community 1's members to community 2's existence, hence community 1's members can now go to community 2 to share their own new hate content. Though this interchange is clearly a complicated human process, the overall takeaway is that a link between two communities (i.e. A and/or B nodes) opens the door to the spread of a new piece of hate content between them -- and the subsequent spread to other  communities that they are linked with (i.e. other A and/or B nodes). Even though the initial link is directed, the two communities can get to know about each other as a result and hence may share new hate content in either direction. Hence we make the links in our model undirected.
Clusters then develop over time that can contain any number of hate (A) and non-hate (B) nodes. Each cluster can grow as isolated nodes link to it and as its own nodes develop links with nodes in other clusters. Opposing this  clustering is the action of platform moderators and hence the fragmentation of a cluster as discussed earlier. The larger the cluster, the more noticeable it will be: hence both coalescence and fragmentation rates will be made proportional to cluster size. This is equivalent to saying that a cluster is chosen for coalescence or fragmentation by first picking any node at random from the entire population of nodes, and then the cluster chosen is the one that this node belongs to. 

Specifically, the numerical simulation proceeds via the following steps: 
\vskip0.3in

\noindent {\bf Step 1 - isolated nodes}: Start with $N$ isolated nodes with the fraction of A (hate) nodes and B (non-hate) nodes given by $p_{A}$ and $p_{B}$.
Hence there are $N$ single-node (i.e. monomer) clusters at the beginning of the simulation.
\vskip0.3in

\noindent {\bf Step 2 - clustering dynamics}:
\begin{itemize}
\item Randomly pick a node $i$ from among all $N$ nodes. The cluster to which this node belongs, is now the cluster of interest for this timestep. A random number generator then decides
which one of three possible events takes place at this timestep: this cluster will fragment, or it will coalesce, or 
nothing happens. 

\item If the randomly picked node $i$ is A-type, then:
\begin{itemize}
\item With probability $\nu_{fA}$, the cluster $G_{i}$ of size $1,2,3,\dots$ that node
$i$ belongs to will fragment, i.e. the intra-cluster links disappear which means that the cluster fragments into isolated nodes.  This is carried out by testing
if the random number $r< \nu_{fA}$.

\item If cluster fragmentation does not occur,  then pick randomly
another node $j$. It will belong to a cluster $G_{j}$ of size $1,2,3,\dots$.
($G_{j}$ will generally be different to $G_{i}$, but it could be the same).  Now we check whether node $j$ is
type A or type B.  If node $j$ is type A (i.e. same type as node
$i$), then the two clusters merge with probability $\nu_{cAA}$.  This is carried out by checking whether
$\nu_{fA} < r < \nu_{fA} + \nu_{cAA}$ (need $\nu_{fA} < r$ to get to this
process).  If no cluster coalescence occurs ($r > \nu_{fA} + \nu_{cAA}$),
then nothing will happen
in this round of the dynamics.  If node $j$ is type B (i.e. different
from node $i$) then the two clusters coalesce with probability
$\nu_{cAB}$.  This is carried out by checking
whether $\nu_{fA}< r < \nu_{fA} + \nu_{cAB}$ (we need $\nu_{fA} < r$ to have 
this process).  If no cluster coalescence occurs ($r > \nu_{fA} + \nu_{cAB}$),
then nothing will
happen in this round of the dynamics.
\end{itemize}

\item If the randomly picked node $i$ is B-type, then:
\begin{itemize}
\item With probability $\nu_{fB}$, the cluster $G_{i}$ that node
$i$ belongs to will fragment.  This is carried out by testing
whether $r< \nu_{fB}$.

\item If cluster fragmentation is not carried out, then pick randomly
another node $j$ that belongs to a cluster $G_{j}$ at the time. 
(As before, $G_{j}$ will generally be different to $G_{i}$, but it could be the same).  Now we need to check whether node $j$ is
type A or type B.  If node $j$ is type B (i.e. same type as node
$i$), then the two clusters coalesce with probability $\nu_{cBB}$.  This is carried out by checking whether
$\nu_{fB} < r < \nu_{fB} + \nu_{cBB}$ (we need $\nu_{fB} < r$ to
have this process).  If no cluster coalescence occurs ($r > \nu_{fB} +
\nu_{cBB}$), then nothing will happen in this round of cluster
dynamics.  If node $j$ is type A (i.e. different from node $i$), then
the two clusters coalesce with probability $\nu_{cBA}$.  This is carried out by checking whether
$\nu_{fB} < r < \nu_{fB} + \nu_{cBA}$ (need $\nu_{fB} < r$ to
get to this process).  If no cluster coalescence occurs ($r > \nu_{fB} +
\nu_{cBA}$), then nothing will happen in this round of cluster
dynamics.
\end{itemize}

\item When a cluster fragments, all the nodes become isolated
nodes (clusters of size 1, i.e. monomers)

\end{itemize}

\vskip0.3in

\noindent {\bf Step 3}: Let Step 2 run for many time steps, until the cluster size distribution settles into a reasonable
steady state.
\vskip0.3in

\noindent {\bf Step 4 - SIR added}: All nodes are in the Susceptible state (S) initially.  At some timestep after the cluster dynamics reach a dynamical equilibrium (i.e. reasonably steady distribution of cluster sizes), select the cluster with the biggest
size at that timestep.  Randomly pick a node in this cluster and change its state into the
Infected state (I).  This means that this node (in-built community) in this cluster now has some new hate content. Unless stated otherwise, this initial infection is placed in a single A (hate) node since these are the nodes that create new hate content. 

\vskip0.3in

\noindent {\bf Step 5 - contagion}: This step consists of one round of the SIR contagion dynamics. Following standard implementations of the SIR process, at each timestep each infected I node (i.e. node (i.e. node holding the new hate content) will
make an attempt to infect (i.e. pass this hate content to) each of the S nodes that it is connected to, i.e. the S nodes in the same cluster as this infected node. This makes sense since each cluster is an interconnected set of nodes, hence they can in principle pass the infection (i.e. this hate content) to each other within the same cluster rather quickly and easily. Moreover, since each node is a single in-built community, it acts as a unit in that once an infection (i.e. new piece of hate content) enters it, all members of this community are infected (i.e. become aware of this hate content) essentially instantaneously on the timescale of a single timestep in the simulation. This mimics the fact that all members of an in-built community have instant access to any new content that gets added to it.

At each timestep in the simulation, each S node will experience infection attempts by all the
I nodes in the same cluster as it. This process is carried
out synchronously for all nodes in all clusters.  
The infection
probability of one infected node to infect an S node to which it is connected (i.e. within the same cluster) is
$\beta$.  After the infection process, every I node in the system (except
those that just got infected) will recover with 
probability $\gamma$.  The homogeneous-mixing assumption within a cluster (i.e. each I node can infect each S node in its cluster with equal probability $\beta$) is a mean-field approximation that becomes more accurate when the cluster's internal structure is well-connected, consistent with the empirical observation that clusters of online communities tend to form dense internal links  \cite{Multi2021,usNature2019,usScience2016}. 

The SIR process is thus parameterized by
$\beta$ and $\gamma$ as shown schematically in Fig. 2. In the present model, $\beta$ and $\gamma$ are independent of node label A and B: every I node infects every neighboring S node in its cluster with the same probability $\beta$ regardless of node label, and every I node recovers with the same probability $\gamma$. This could be generalized in future work to A or B dependent infection and recovery rates, but such a generalized model would inherit all the complexities of the present case plus additional ones --- hence it is essential to understand the species-independent version first. We also note that the species-dependent infection case is already partially explored through the digital vaccination mechanism (Case~III), where B nodes have a reduced infection rate $\chi = \phi\beta < \beta$. In addition, the model already allows species-dependent fragmentation probabilities ($\nu_{fA} \neq \nu_{fB}$), which captures the scenario where clusters containing more hate nodes may be fragmented with higher probability.

\vskip0.3in

\noindent {\bf Step 6}: Repeat Step 2 for the clustering dynamics $m$ times. Unless stated otherwise, we use $m = N$ as the default value throughout this paper, which corresponds to one Monte Carlo step in which every node is on average given a chance to participate in the clustering dynamics. To explain why $m$ matters, consider the case of $m=1$: this means that one update of the SIR process takes place for every I node and
affects all the S nodes in the system, but only one node is picked
for the clustering dynamics.  Hence the clustering dynamics for $m=1$ is slow compared with the
SIR dynamics, and importantly 
it leads to very low mixing of
the nodes.  Having a more general value of the parameter $m$
here allows $m\geq 1$ rounds of the clustering dynamics to take place before one
round of SIR dynamics is implemented.  For our system
with $N$ nodes, using $m=N$ is like one Monte Carlo step
(c.f. spin-flip attempts in Ising model simulations so that every
spin is on average given a chance to flip). Figure~\ref{fg:mcheck} shows how $m$ affects the infection profile and hence spreading. The interleaving of clustering and SIR dynamics for our choice of $m=N$ means that the network topology continues to evolve at scale while content is actively spreading, capturing the empirical reality that the coalescence-fragmentation and contagion co-exist.

\vskip0.3in

\noindent {\bf Step 7}: Repeat Step 5 and Step 6 for many time steps, until all I nodes
disappear (i.e. $I(t) = 0$, meaning they have all transitioned to the R state in SIR). This means that the system's infection is now over. The quantity $R/N$ reported throughout this paper is the final fraction of recovered (R) nodes at termination, representing the total fraction of all $N$ nodes that were ever infected during the SIR process. In every simulation run, the number of R nodes eventually reaches a maximum value corresponding to the disappearance of all I nodes, and $R/N$ does not change from then on: it is this value $R/N$ that appears in all our plots. 

\vskip0.3in

Using this numerical simulation, we can establish whether a new piece of hate content with a given $\beta$ and $\gamma$ will spread system-wide or not as a result of this dynamical linking within and between A (hate) and B (non-hate) communities (nodes). In principle, using the empirical data of Refs. \cite{Multi2021,Lupu,npjMin}, specific values of $\beta$ and $\gamma$ can be inferred for  particular pieces of hate content -- but we will leave these as parameters in our analysis for generality. In Sec. III we explore this same coalescence-fragmentation--plus--SIR model analytically by deriving approximate mean-field equations. The numerical and analytic results are presented and compared in Sec. IV. 

\section{Mean-Field Descriptions of the Model}

\label{sec:theory}
 
In this section we derive mean-field (i.e. effective medium) equations for comparison with the numerical model. These equations offer quantitative insight into how the system's parameters need to be controlled to prevent or generate system-wide spreading from a non-spreading state. We note that our analytic analysis in this section has broader potential applications than the present study, since the node species labels A and B  could in the future be used to represent other node properties instead of a hate node and a non-hate node. 

\subsection{Two-species clustering dynamics}
\label{sec:P}
 
We start by providing a theoretical description of the effects of the two-species dynamical clustering {\em without} the SIR contagion process. 
Specifically, Eq. 1 below  generalizes the single-species Master equation from Ref. \cite{2025book} to the A and B species situation in Fig. 2. It represents the key feature of the cluster dynamics: nodes are picked randomly -- and hence the clusters they belong to  are picked with a probability proportional to their size -- and then pre-assigned probability parameters determine whether these clusters coalesce or fragment. Because  our  description is at the cluster level, we do not need to keep track of exactly which links are in which cluster, or how many links are in each cluster. For example, though a cluster of say 4 nodes has a range of possible internal numbers of links, all that matters is that the 4 nodes are interconnected and hence form a cluster. This also makes sense in the real world,  because any two social media communities in the same cluster will want to interact and hence form a link to each other, even if they had no direct link when they initially joined the cluster: indeed, it is observed empirically that clusters have dense internal links and hence it is reasonable to regard them as being approximately fully connected. 

Equation 1 is a Master equation for the time evolution of $P_{ij}(t)$ which is the probability that two nodes $i$ and $j$ selected randomly and independently from across the  system at time $t$, belong to the same cluster of nodes. This quantity $P_{ij}(t)$ will play a key role for our theory in the presence of the content (i.e. virus) spreading, since the passing of a virus from one node to another depends on there being a link present between them. Taking the node species label $\alpha$ as implicit in $i$ and $j$ in $P_{ij}(t)$, the two-species (A and B) Master equation for $P_{ij}(t)$ becomes:

\begin{eqnarray}
\label{eq:dp}
\frac{{\partial} P_{ij}}{{\partial} t}  =   -P_{ij} \frac{1}{N} \sum_{k \in \{...i...\}}  P_{ki} [\delta_{kA} \nu_{fA} + \delta_{kB} \nu_{fB} ] \ \ \ \ \ \ \ \  \\
 +   [1- P_{ij}]  \f{1}{N} \sum_{m  \in \{...i...\}}  P_{mi} \frac{1}{N} \sum_{n  \notin \{...i...\}}  P_{nj} 
[\delta_{mA}\delta_{nA}\nu_{cAA}\ \ \ \nonumber \\ +\delta_{mB}\delta_{nB}\nu_{cBB}+
\delta_{mA}\delta_{nB}\nu_{cAB}+ \delta_{mB}\delta_{nA}\nu_{cBA}]  \nonumber 
\end{eqnarray}

\noindent with sums over nodes either within a cluster or outside a cluster, and coalescence and fragmentation probabilities per timestep $\{\nu_{c{\alpha}{\alpha'}}\}$ and $\{\nu_{f{\alpha}}\}$ as in Fig. 2. We allow the coalescence probabilities to be different when A-A, A-B (B-A), or B-B nodes are picked for coalescence (i.e., $\nu_{cAA}, \nu_{cAB}, \nu_{cBB}$) since hate communities and non-hate communities will have different characteristics. Likewise the fragmentation probabilities are $\nu_{fB}$  and $\nu_{fA}$. 

To explain Eq. 1, its right-hand side consists of two separate terms corresponding to two separate situations that
 can change the value of $P_{ij}(t)$ at each timestep. The first situation (i.e. the first term) is where
the nodes $i$ and $j$  are in the same cluster and this cluster fragments, e.g. due to moderator shutdown. The second situation (i.e. the second term) is where the nodes $i$ and
$j$ are not currently in the same cluster, but the clusters that they belong to coalesce because a new (hyper)link gets created between two of these clusters' nodes. As a result of Eq. 1's temporal evolution, a cluster will typically include some mix of both A and B nodes.
Going into the specifics, the first line of Eq. 1 corresponds to the probability that nodes $i$ and $j$ belong to the same cluster (i.e. factor $P_{ij}$), and a node $k$ that is randomly selected (i.e. factor $\frac{1}{N}$) also belongs to the same cluster as $i$ and $j$ (i.e. factor $P_{ki}$): with a certain probability that depends on the species label (A or B), this cluster then fragments. The second and third lines of the equation depict the case where nodes $i$ and $j$ do not belong to the same cluster (i.e. factor $1-P_{ij}$). One node $m$ is chosen (i.e. factor $\frac{1}{N}$), which is in the same cluster as $i$ (i.e. factor $P_{mi}$), and one node $n$ is picked (i.e. factor $\frac{1}{N}$) from the same cluster as $j$ (i.e. factor $P_{nj}$). They coalesce according to the two species labels. $N$ is the total system node number. Hence Eq. 1 captures the physics of new links causing clusters of nodes to coalesce and hence grow, but where each such cluster may suddenly fragment if targeted by moderators \cite{usNature2019,PRL2023}. The use of a product kernel in Eq. \ref{eq:dp} can be generalized, but prior work has shown that this product kernel form is indeed consistent with empirical online communications data \cite{Palla07,PRL2023,usScience2016}. 

We now treat Eq. 1 in a mean-field way. Specifically, we take an average over all $i$ and $j$ pairs, so $P_{ij}\rightarrow P$. We replace sums like $\sum_{k\in\{..i.\}}P_{ki}$ by a mean value $C/N$ where $C$ is the average cluster size, and sums like $\sum_{n\not\in\{..i.\}}P_{nj}$ by $(1-C/N)$. The probability with which A or B labels  occur when evaluating Eq. 1 in this average way, is equal to their relative proportion $p_A$ or $p_B$; and hence
averaged coalescence and fragmentation probabilities emerge given by 
\begin{equation}
\tilde\nu_c = {\nu_{cAA} p_A^2 + \nu_{cBB} p_B^2 + \nu_{cAB} p_A p_B+ \nu_{cBA} p_B p_A} \nonumber
\end{equation}
\begin{equation}
\tilde\nu_f = p_A \nu_{fA} + p_B \nu_{fB} \ .
\end{equation}
We set $\nu_{cAB}=\nu_{cBA}$ from now on for simplicity. 
This averaging gives the following mean-field version of Eq. 1: 

\begin{equation}\label{master2}
\small{\dfrac{\mathrm{d} P }{\mathrm{d}t} = -P . \frac{1}{N} . \frac{C}{N} . \tilde{\nu}_f+(1-P) . \frac{1}{N} . \frac{C}{N} . \frac{1}{N} . \bigg(1-\frac{C}{N}\bigg) . 
\tilde{\nu}_c} 
\end{equation}
We next consider the steady-state limit, $\frac{\mathrm{d} P }{\mathrm{d}t}=0$. 
Taking also the limit that the average cluster size $C\ll N$, this steady-state mean-field version of Eq. 1 (i.e. Eq. 3) therefore yields
\begin{equation}\label{master3}
 {P  N   \tilde{\nu}_f = (1-P)  \tilde{\nu}_c\ \ }\ .
\end{equation}
Solving Eq. 4 for $P$ yields
\begin{equation}
\label{eq:P1}
 P = {\tilde \nu _c}({\tilde \nu _c  + N\tilde \nu _f })^{-1}\ .
 \end{equation} 
 \noindent 
This Eq. 5 actually holds for any number of species $M$ using
\begin{equation}
\label{eq:equation3}
\tilde \nu _c = \sum_{\alpha=1}^M \nu _{c\alpha\alpha} p_\alpha p_\alpha+\sum_{\alpha\neq {\alpha'},1}^M \nu _{c\alpha{\alpha'}} p_\alpha p_{\alpha'},\ \ \ \tilde \nu _f  = \sum_{\alpha=1}^M \nu _{f\alpha} p_\alpha
\end{equation}
with $p_\alpha$ and $p_\alpha'$ being the fraction  of nodes of species $\alpha$ and $\alpha'$, while $\nu _{c\alpha\alpha}$, $\nu _{c\alpha\alpha'}$ and $\nu _{f\alpha}$ are the respective coalescence and fragmentation probabilities.
In the approximate limit that $N {\tilde \nu _f} \gg \tilde{\nu_c} $, Eq. 5 yields the following for any number of species $M$ of nodes:
\begin{equation}
P = \f{\tilde{\nu}_c}{N \tilde{\nu}_f} \ . \label{eq:P}
\end{equation}

\subsection{Adding the SIR process}

We now add in the SIR contagion process for each node, recalling our simplification that each node is either in the Susceptible (S), Infected (I) or Recovered (R) state, where `infected' means that the node (i.e. A or B community) contains the new piece of hate content and that all its members have been exposed to it. As in the numerical simulation described in Sec. II, for infected nodes that are part of a cluster at any given time, any given I node will infect (i.e. share the content with) any given S node in its cluster with probability $\beta$ per timestep, and every I node recovers (i.e. becomes an R) with probability $\gamma$. More complex processes than SIR could also be treated by inserting the appropriate compartments into the equations that follow in this section.

\subsection{Effective Medium Theory (EMT) and beyond (BEMT)}

Consider one A node getting infected at some point in time after the coalescence-fragmentation clustering has reached its dynamical steady state: we call this time $t=0$. We will now develop equations for the subsequent contagion in the system at two levels of approximation -- Effective Medium Theory (EMT) and Beyond Effective Medium Theory (BEMT) -- where the two types of averaging and their names are inspired by traditional physics modeling of transport in alloys. The BEMT is more fine-grained than the EMT, but it is not a priori obvious which one will give the best quantitative fit to the numerical simulations for a given set of parameter values. Seeing which is closer for a given parameter range informs the best way to do the averaging for that parameter range, and hence the best way to picture the complex coalescence-fragmentation-SIR dynamical system for that parameter range.

\vskip0.2in

\noindent (1) {\bf Effective Medium Theory (EMT)}. We start from the simplest version of the SIR viral process, which is to use $P \cdot S \cdot I$ for the number of S-I node contacts. The dynamical equations can then be written as:\begin{eqnarray}
 \frac{{dS}}{{dt}}&=&  - \beta PSI, \label{eq:dS}\\
 \frac{{dI}}{{dt}}&=&  \beta PSI - \gamma I, \label{eq:dI}\\
 \frac{{dR}}{{dt}}&=& \gamma I \label{eq:dR}
\end{eqnarray}

At the Effective Medium Theory (EMT) level of approximation, we simply insert the earlier expressions for $P$ into these SIR equations governing the entire subpopulations of S, I, and R. In other words, the EMT averages within and across species which, crudely speaking, will tend to underestimate the impact of correlations. For a multi-platform system of $N$ nodes, with $S_0$ susceptible nodes at time $t=0$ when the content first appears in one or more nodes, the threshold condition for no system-wide spreading of this content can be derived from Eq. 9 by requiring no initial growth in $I$, i.e. $ \frac{{dI}}{{dt}}\leq 0$ at time $t=0$. Hence $1\geq PS_0\beta /\gamma\equiv {\kappa}^{\rm EMT}$ from Eq. 9. This means that if ${\kappa}^{\rm EMT}\leq 1$, there is no system-wide spreading. For this reason, ${\kappa}^{\rm EMT}$  is referred to as the basic reproduction rate. If there is only one I node at $t=0$, then $S_0=N-1$ and ${\kappa}^{\rm EMT}=(N-1)P\beta /\gamma$. Hence, the EMT predicts that system-wide spreading is prevented when
\begin{equation}
\label{eq:kappa1}
{\kappa}^{\rm EMT}\equiv (N-1) {\tilde \nu _c} \beta [({\tilde \nu _c  + N\tilde \nu _f })\gamma]^{-1} < 1\ .
\end{equation}

\noindent Equation \ref{eq:kappa1} shows explicitly the interplay, and hence trade-offs for controlling spreading and also intervention, between the node clustering dynamics (${\tilde \nu _c}$ and ${\tilde \nu _f}$) and the viral dynamics ($\beta$ and $\gamma$), or equivalently their respective timescales given by the reciprocals. 

A key implication of this threshold expression in Eq. 11, is that even if the condition $\beta/\gamma > 1$ holds (which would mean that system-wide spreading would occur in the usual ideal SIR system), the no-spreading condition in Eq. \ref{eq:kappa1} can still be satisfied if ${\tilde \nu _c}$ is sufficiently small and/or
${\tilde \nu _f}$ is sufficiently large. 

\vskip0.1in
This result shows explicitly for policymakers that {\em the link dynamics between nodes of the same type (e.g. hate-hate) and nodes of different types (hate--non-hate) in Fig. 2 can be jointly manipulated to prevent system-wide spreading, even if the implicit contagiousness of the content is sufficiently high for system-wide spreading to occur} (i.e. $\beta/\gamma>1$). 
At the same time, Eq. 11 warns that the real-world reality of so much hate content now spreading system-wide online, means that current ${\tilde \nu _c}$ and ${\tilde \nu _f}$ values need mitigating. Equation 11 also raises doubts about mitigation strategies that focus solely on the implicit contagiousness ($\beta/\gamma$) of a specific piece of new hate content.

\vskip0.1in
At this rather simple level of mean-field approximation, it is relatively easy to add the effect of online digital vaccination, i.e. we assume the non-hate communities B are made more resistant  to certain content. This might in principle happen naturally because of their nature, or by preemptively providing them with real facts. This means that B nodes have a lower infection rate $\chi$. The dynamical equations then become
\begin{eqnarray}
 \frac{{dS_A }}{{dt}} &=&  - \beta PS_A I, \\
 \frac{{dS_B }}{{dt}} &=&  - \chi PS_B I, \\
 \frac{{dI}}{{dt}} &=& P(\beta S_A  + \chi S_B )I - \gamma I, \\
 \frac{{dR}}{{dt}} &=& \gamma I .
\end{eqnarray}
We note that all the above equations can be regarded as a
site-level approximation in physics terminology, with an average probability that two
sites are connected, i.e. not distinguishing the links connecting
the same and different species of nodes. 

\vskip0.1in

\noindent (2) {\bf Beyond Effective Medium Theory (BEMT)}. We now go beyond the above EMT by separating the links into A-A, A-B (B-A), B-B types, i.e. we break apart $P$ into its species-specific components (see Appendix A) and then we insert these species-specific contributions into the coupled SIR equations for A and B separately. Since this BEMT approach only averages within a species, it may crudely speaking overestimate the impact of correlations. The BEMT leads to more detailed dynamical equations, for which the general Eqs. \ref{eq:dS}, \ref{eq:dI} and \ref{eq:dR} are now expressed in terms of the partial contributions of A-A, B-B and A-N (B-A). The equations for BEMT become:
\begin{eqnarray}
 \frac{{dS_A }}{{dt}} &=&  - \beta LP_{AA} \frac{{S_A I_A }}{{L_A }} - \beta LP_{AB} \frac{{S_A I_B }}{{L_{AB} }}, \label{eq:dSH} \\
 \frac{{dI_A }}{{dt}} &=& \beta LP_{AA} \frac{{S_A I_A }}{{L_A }} + \beta LP_{AB} \frac{{S_A I_B }}{{L_{AB} }} - \gamma I_A, \label{eq:dIH}  \\
 \frac{{dR_A }}{{dt}} &=& \gamma I_A,\label{dRH}  
\end{eqnarray}
\begin{eqnarray}
   \frac{{dS_B }}{{dt}} &=&  - \beta LP_{BB} \frac{{S_B I_B }}{{L_B}} - \beta LP_{BA} \frac{{S_B I_A }}{{L_{BA} }}, \label{dSN} \\
   \frac{{dI_B }}{{dt}} &=& \beta LP_{BB} \frac{{S_B I_B }}{{L_B}} + \beta LP_{BA} \frac{{S_B I_A }}{{L_{BA} }} - \gamma I_B, \label{dIN}  \\
   \frac{{dR_B }}{{dt}} &=& \gamma I_B, \label{dRN}
\end{eqnarray}
where the  $L$-quantities calculate the number of possible links for different levels of partitioning of the systems' nodes: for the total system $L=N(N-1)/2$, and then the respective partial contributions are $L_A=N_A(N_A-1)/2$, $L_{AB}=L_{BA}=N_A N_B$,
and $L_B=N_B(N_B-1)/2$ where $N_A$ and $N_B$ are the number of A and B nodes. Equations \ref{eq:meanPHH}-\ref{eq:meanPNN} in the Appendix
provide expressions for $P_{AA}$, $P_{AB}(\equiv P_{BA})$, and $P_{BB}$ within this BEMT theory.

Similarly to the EMT, we can try to determine the non-spreading condition -- though this will reveal a new complication that we then overcome. We start by adding Eqs. \ref{eq:dIH} and \ref{dIN}, to obtain a condition for which $dI/dt =0$ at a given time $t$.  This condition for $dI/dt = 0$ at a given time $t$ is 

\begin{equation}
\begin{split}
\frac{\beta }{\gamma}[ \frac{{(N - 1)P_{AA} }}{{(Np_A  -
1)p_A }}\frac{{S_A I_A }}{I} + \frac{{(N - 1)P_{AB} }}{{2Np_A p_B
}}(\frac{{S_A I_B }}{I} + \frac{{S_B I_A }}{I}) \\ + \frac{{(N -
1)P_{BB} }}{{(Np_B  - 1)p_B }}\frac{{S_B I_B }}{I}] = 1
\;, \label{eq:dIdtequals0}
\end{split}
\end{equation}

\noindent If Eq. \ref{eq:dIdtequals0} is satisfied, $I$ will not change in
the next time step. This hence suggests the following generalized expression for the instantaneous reproduction rate
$\kappa$ at time $t$, which involves the parameters and the instantaneous numbers
of S, I and R nodes:

\begin{equation}
\begin{split}
{\kappa}^{\rm BEMT} = \frac{\beta}{\gamma}[ \frac{{(N - 1)P_{AA}
}}{{(Np_A - 1)p_A }}\frac{{S_A I_A }}{I} + \frac{{(N - 1)P_{AB}
}}{{2Np_A p_B }}(\frac{{S_A I_B }}{I} + \frac{{S_B I_A }}{I}) \\ +
\frac{{(N - 1)P_{BB} }}{{(Np_B  - 1)p_B }}\frac{{S_B I_B }}{I}
]. \label{eq:kappa03}
\end{split}
\end{equation}

\noindent The new complication compared to EMT is that ${\kappa}^{\rm BEMT}$  will take on different values at different times.
In any case, ${\kappa}^{\rm BEMT} > 1$ (${\kappa}^{\rm BEMT} < 1$) indicates an increase (a
drop) in $I(t) = I_{A}(t) + I_{B}(t)$ in the next time step. We stress
that $S_{A}$ and $S_B$ are the instantaneous numbers, 
not the initial values. Hence 
the above BEMT analysis predicts that the spreading in a population with coupled dynamic clusters of nodes, will depend
on the values of this instantaneous ${\kappa}^{\rm BEMT}$ as the system evolves.
This is in sharp contrast to the case of a {\em single} well-mixed
population, in which an initial drop in the number of infected
nodes guarantees that the number will not increase in time, i.e. the initial value of the reproduction rate (which we denote as ${\kappa}^{\rm BEMT}_0$ and which involves the initial number of
susceptible nodes) sets the spreading threshold.  In the 
case of multiple species, we have coupled clusters of nodes of different species and hence even if the number of infected nodes
drops initially in one cluster, members of the other clusters may be
infected through the coupling so that the total number of infected
nodes grows in later time steps and spreading is still
possible. 

If we focus on the initial reproduction rate (i.e. ${\kappa}^{\rm BEMT}_0$) and we suppose that all I nodes that exist at this time are A type (i.e. hate), we get from Eq. 23 that
\begin{equation}
{\kappa}^{\rm BEMT}_0 =\frac{\beta }{\gamma
}\left[\frac{{(N-1)}}{{p_A(Np_A-1)}}P_{AA}S_{A,0}+\frac{{(N-1)}}{{2Np_B
p_A }}P_{AB} S_{B,0}\right] \label{eq:kappa}
\end{equation}
where $S_{A,0}$ and $S_{B,0}$ are $S_{A}$ and $S_B$ at time $t=0$. If ${\kappa}^{\rm BEMT}_0 > 1$, there will be an initial spreading.  For the
particular initial condition of only one I node with this node of A type, which we use in most of our simulations,  then
\begin{equation}
{\kappa}^{\rm BEMT}_0
=\frac{\beta}{\gamma}\frac{{(N-1)}}{{p_A}}\left[{P_{AA}+\frac{P_{AB}}{2}}\right]
\end{equation}
 which means that the initial system-wide spreading is prevented when

{\small\begin{equation}
\label{eq:kappa02}
{\kappa}^{\rm BEMT}\equiv \frac{{(N-1) P}}{{p_A \tilde \nu _c}}
\big[\nu_{cAA} p_A^2 +\frac{1}{2}(\nu _{cAB}  + \nu_{cBA} )p_B p_A
\big]\frac{\beta}{\gamma} < 1
\end{equation}}

\noindent which is the BEMT version of Eq. 11.
Hence if we then take $p_A=p_B=1/2$, we get the EMT result $(N -
1)P\beta/\gamma$ as expected since the effect of species heterogeneity is effectively then removed. 
Equation \ref{eq:kappa02}  predicts that system-wide spreading can be prevented even if $\beta/\gamma>1$, as found also for the EMT case. 

For the case of online digital vaccination, the dynamical equations are now given by
\begin{eqnarray}
 \frac{{dS_A }}{{dt}} &=&  - \beta LP_{AA} \frac{{S_A I_A }}{{L_A }} - \beta LP_{AB} \frac{{S_A I_B }}{{L_{AB} }}, \\
 \frac{{dI_A }}{{dt}} &=& \beta LP_{AA} \frac{{S_A I_A }}{{L_A }} + \beta LP_{AB} \frac{{S_A I_B }}{{L_{AB} }} - \gamma I_A,  \\
 \frac{{dR_A }}{{dt}} &=& \gamma I_A,  \\
 \frac{{dS_B }}{{dt}} &=&  - \chi LP_{BB} \frac{{S_B I_B }}{{L_\mathrm{n} }} - \chi LP_{BA} \frac{{S_B I_A }}{{L_{BA} }}, \\
 \frac{{dI_B }}{{dt}} &=& \chi LP_{BB} \frac{{S_B I_B }}{{L_\mathrm{n} }} + \chi LP_{BA} \frac{{S_B I_A }}{{L_{BA} }} - \gamma I_B,  \\
 \frac{{dR_B }}{{dt}} &=& \gamma I_B.  \\
 \nonumber
\end{eqnarray}

\vs

\vs

Both the EMT and BEMT mirror the full simulations and the real-world reality in predicting that {\em although hate content can appear isolated and largely eradicated among B nodes (i.e. non-hate communities), it can simultaneously move through inter-cluster links to the A nodes (i.e. hate communities) where it revives before later re-emerging among B nodes B}. This can in turn give rise to re-entrant spreading phase behavior, as shown in Secs. IV and V.

\section{Numerical Simulations vs. Analytic Approximations (EMT, BEMT)}

\label{sec:simulation}

\begin{figure}[!t]
\centering
\epsfig{figure=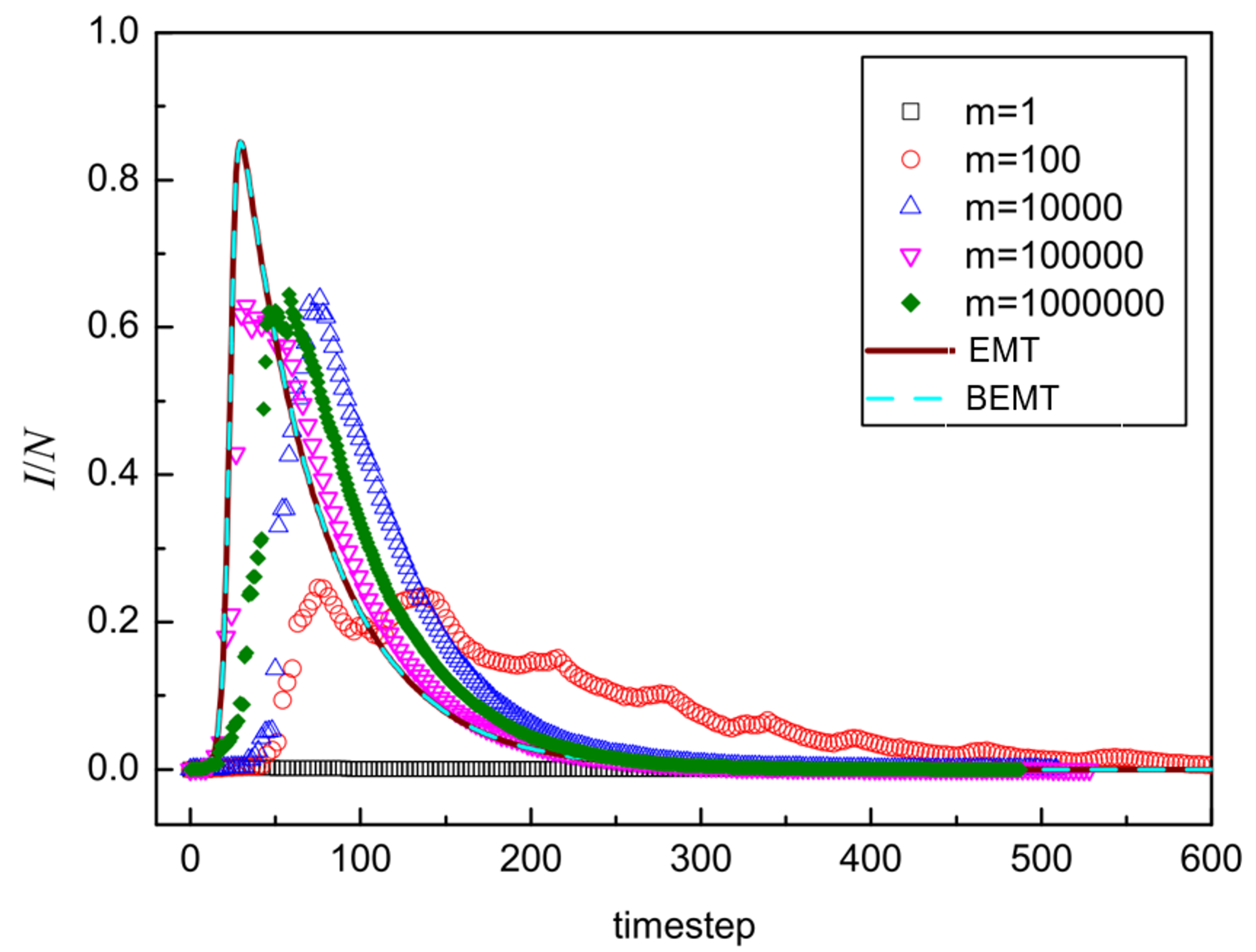,width=\linewidth} \caption{Vertical scale shows the
fraction of infected nodes I as a function of the simulation timesteps for Case I. The
parameters are: $N=10000$, $p_A=p_B=0.5$, $\nu_c=0.5$,
$\nu_f=0.005$, $F=1.0$, $\beta=0.005$, and $\gamma=0.02$. Results
of EMT and BEMT are also included for comparison: they match the simulation output very well in the limit of large $m$.}
\label{fg:mcheck}

\end{figure}

Showing comparisons between the EMT, BEMT  and the numerical simulation in a concise way is complicated: the large number of parameters even with just two species (A and B), mean that we must here consider only a subspace of this parameter space. This is already complicated enough, as our extensive results will now show. We will therefore do this by focusing on the following  particular cases:

\noindent \textbf{Case I} : In this case, we choose the two node species A and B to have the same intra-coalescence probability ($ \nu $)  while the inter-coalescence probability is a fraction $F$ of it ($ F  \nu$). 
In summary, we choose 
$\nu_{cAA}=\nu_{cBB}=\nu_c$, $\nu_{cAB}=\nu_{cBA} = F \nu_c$ and also $\nu_{fA}=\nu_{fB}=\nu_f$. Figure ~\ref{fg:mcheck} shows the results for different values of the coalescence-fragmentation steps ($m$) for Case I.  The key point is that the EMT and BEMT become more accurate for larger values of $m$. We take $N=m$ in subsequent simulations as mentioned before.

\noindent \textbf{Case II: } In this case, we choose the coalescence probability involving one node species  to be a fraction of the intra-coalescence probability of the other. In summary, we choose
$\nu_{cAA}=\nu_c$, $\nu_{cAB}=\nu_{cBA}=\nu_{cBB}=G\nu_c$ and also $\nu_{fA}=\nu_{fB}=\nu_f$.

\noindent Case I allows us to explore the effect of the inter-coalescence probability in relation to variations in the contact ratio. Case II allows us to explore the effect of coalescence involving a single species type as the contact ratio fluctuates.  We will also study Case III which allows us to assess the impact of vaccination on population immunity, specifically focusing on the fraction of nodes in the population that have been vaccinated.

\begin{figure}[!t]
\centering
\epsfig{figure=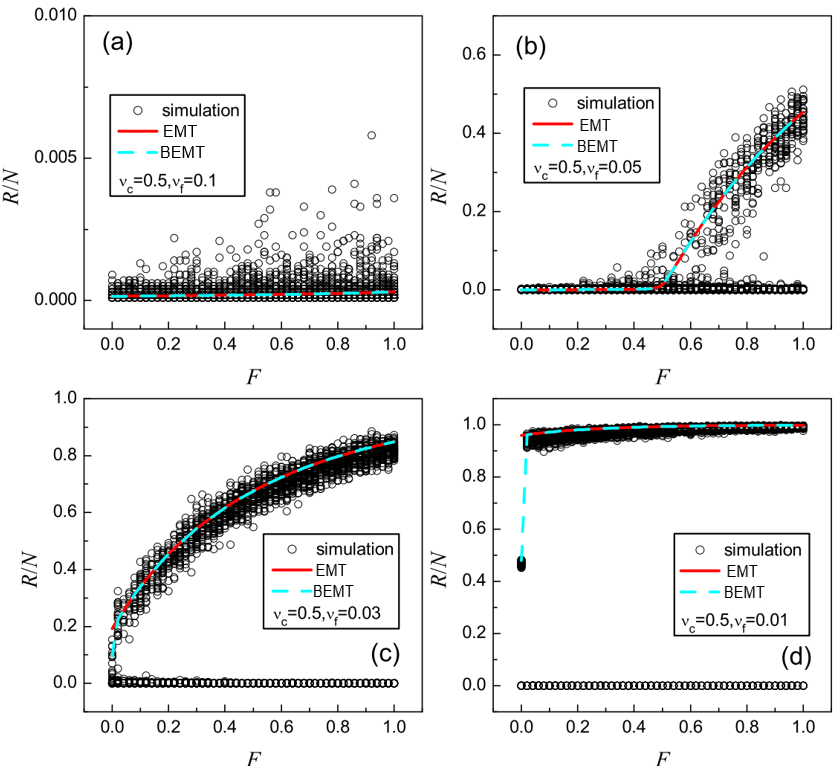,width=\linewidth} \caption{Simulation results for Case I based on different fragmentation rates. The diagrams show the fraction of nodes that reach state R as a function of $F$ for different values of $\nu_f$, i.e., $\nu_f = 0.1, 0.05, 0.03, 0.01$ for (a), (b), (c), (d). A significant non-zero value can be interpreted as system-wide spreading since it means that a significant fraction of all nodes have reached the R state and hence have been infected. The theory curves are for EMT and BEMT as described in the theory section. The parameters are: $N=10000$, $m=N$, $p_A=p_B=0.5$, $\nu_c=0.5$, $\beta=0.002$, and $\gamma=0.015$. The symbols are obtained from the numerical simulation of the model.  For each value of $F$, there are 100
trials.} \label{fg:RvsFcaseIa}

\end{figure}

\begin{figure}
\centering
\epsfig{figure=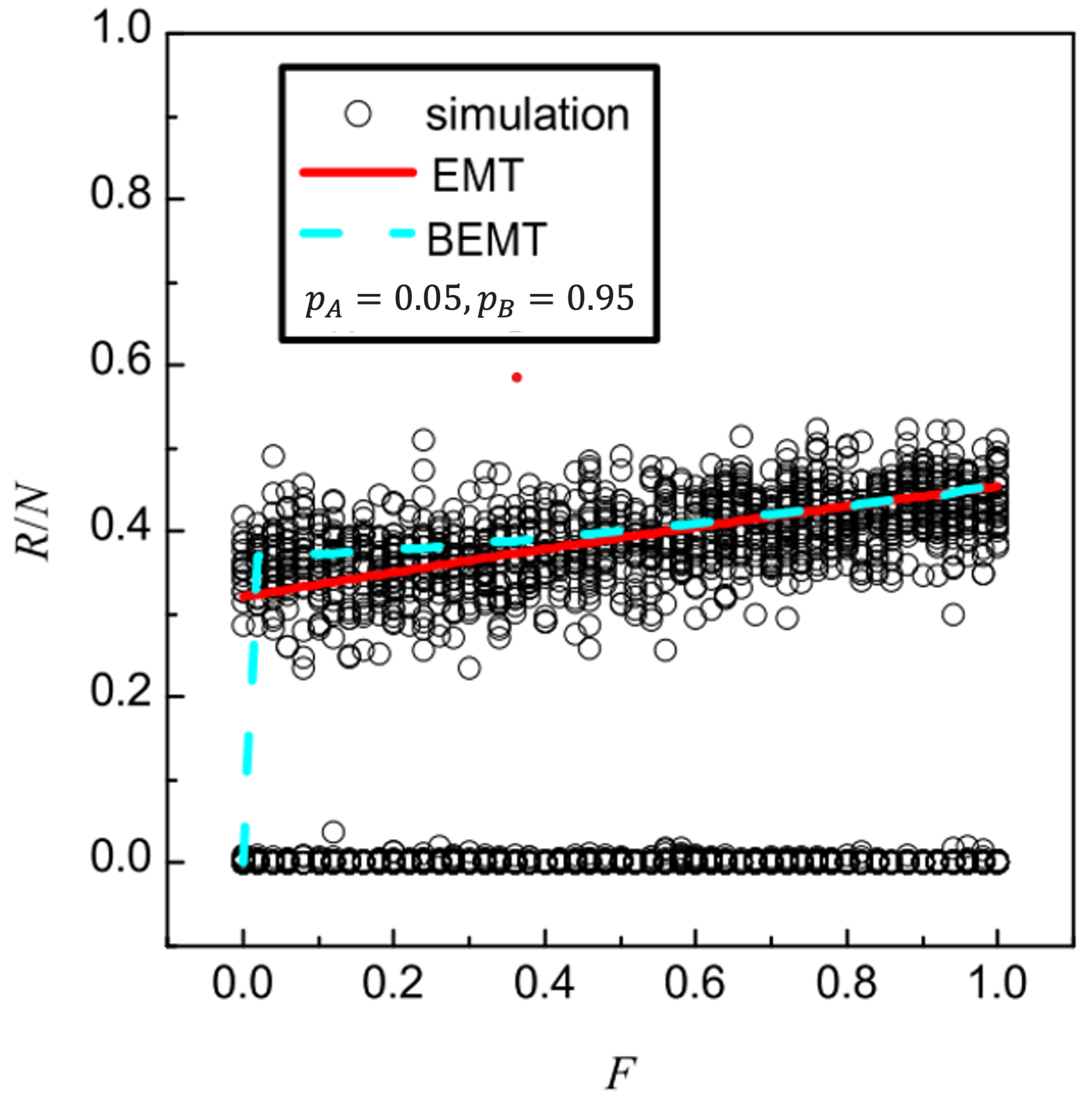,width=0.8\linewidth} \caption{The
fraction of nodes reaching state R as a function of $F$. The red solid curve and blue dash curve are for EMT and BEMT as defined in the theory section respectively. The parameters are: $N=10000$, $m=N$, $p_A=0.05$, $p_B=0.95$, $\nu_c=0.5$, $\nu_f=0.05$,
$\beta=0.002$, and $\gamma=0.015$. The symbols are obtained by
simulations.  For each value of $F$, there are 100 trials.}
\label{fg:RvsFcaseIb}

\end{figure}

\subsection{Case I  (Fig.~\ref{fg:RvsFcaseIa} and Fig.~\ref{fg:RvsFcaseIb})}

In Fig.~\ref{fg:RvsFcaseIa}, we can see that the system may (or may not) show
an outbreak (i.e. system-wide spreading) in 100 trials for each value of $F$. A significant non-zero value of $R/N$ can be interpreted as system-wide spreading since it means that a significant fraction of all nodes have reached the R state and hence have been infected. The
fraction of runs showing spreading (and hence the probability of spreading) depends on the model parameters.  For the
realizations with spreading, the $R/N$ behavior is very well described by both theories EMT and BEMT. The theoretical results for EMT and BEMT are very similar since the system is symmetric with respect to species fractions and SIR parameters. It should be noted that in Fig.~\ref{fg:RvsFcaseIa}(d), EMT cannot predict the $F=0$ case very well. The reason is that the connection between the two species is completely broken but EMT still mixes them in the SIR equations. Figure~\ref{fg:RvsFcaseIb} shows a case where the species fractions are different.  In this case, the EMT and BEMT results differ more.

\begin{figure}[!t]
\centering
\epsfig{figure=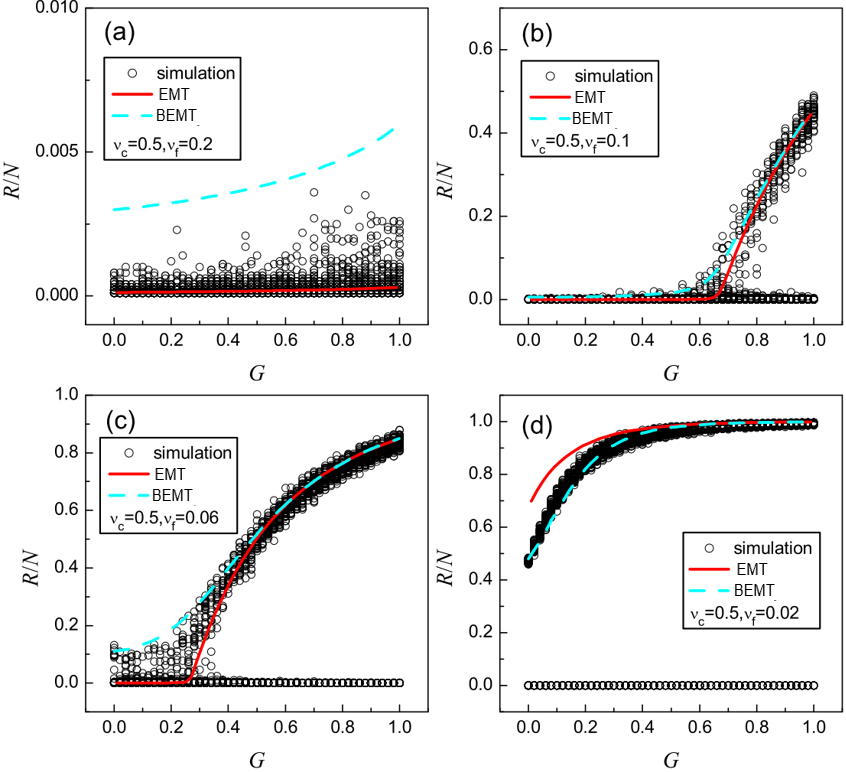,width=\linewidth} \caption{Simulation results for Case II based on different fragmentation rates. The diagrams illustrate the
fraction of nodes reaching state R as a function of $G$ for different values of
$\nu_f$, i.e. $\nu_f = 0.2, 0.1, 0.06, 0.02$ for (a), (b), (c), (d). The red solid curve and blue dash curve are for EMT and BEMT as defined in the theory section respectively. The parameters are: $N=10000$, $m=N$, $p_A=p_B=0.5$,
$\nu_c=0.5$, $\beta=0.004$, and $\gamma=0.015$. The symbols are
obtained by simulations, with 100 trials for each value of $G$.}
\label{fg:RvsGcaseII}

\end{figure}

\subsection{Case II (Fig.~\ref{fg:RvsGcaseII})}

In this case, the cluster process is asymmetric with respect to the two species types A and B. In Fig.~\ref{fg:RvsGcaseII}, the system may (or may not) show spreading across 100 trials. For the realizations with spreading, BEMT generally works better than EMT. For example, in scenarios characterized by a low fragmentation rate (Fig.~\ref{fg:RvsGcaseII} (d)), the EMT suggests a consistently elevated system-wide spreading rate across all values of $G$ when compared to simulation outcomes. In contrast, the BEMT aligns more closely with the patterns observed in the simulation results.

\begin{figure}[!t]
\centering
\epsfig{figure=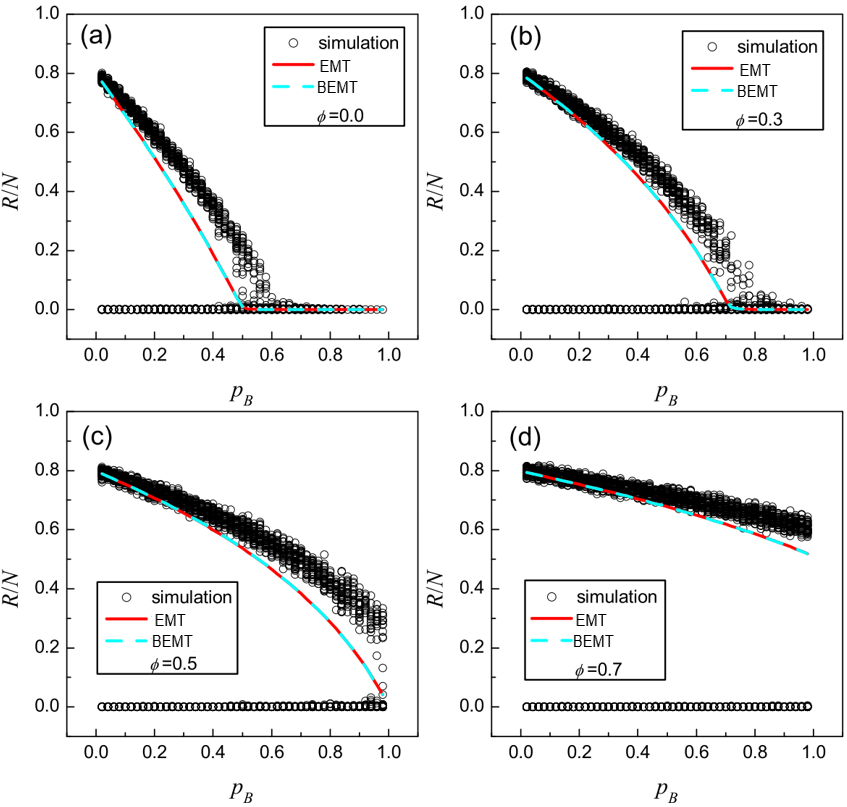,width=\linewidth} \caption{ Simulation results for Case I with an immunity effect for the species B. The diagrams describe the fraction of nodes reaching state R as a function of $p_B$ for different immunization coefficient $\phi$ ($\phi = 0, 0.3, 0.5, 0.7$ for diagram (a), (b), (c), (d).) The red solid curve and blue dash curve are for EMT and BEMT as defined in the theory section respectively. The parameters are: $N=10000$, $m=N$,
$\nu_c=0.02$, $\nu_f=0.01$, $\beta=0.01$, and $\gamma=0.01$.}
\label{fg:RvspBF}

\end{figure}

\begin{figure}[!t]
\centering
\epsfig{figure=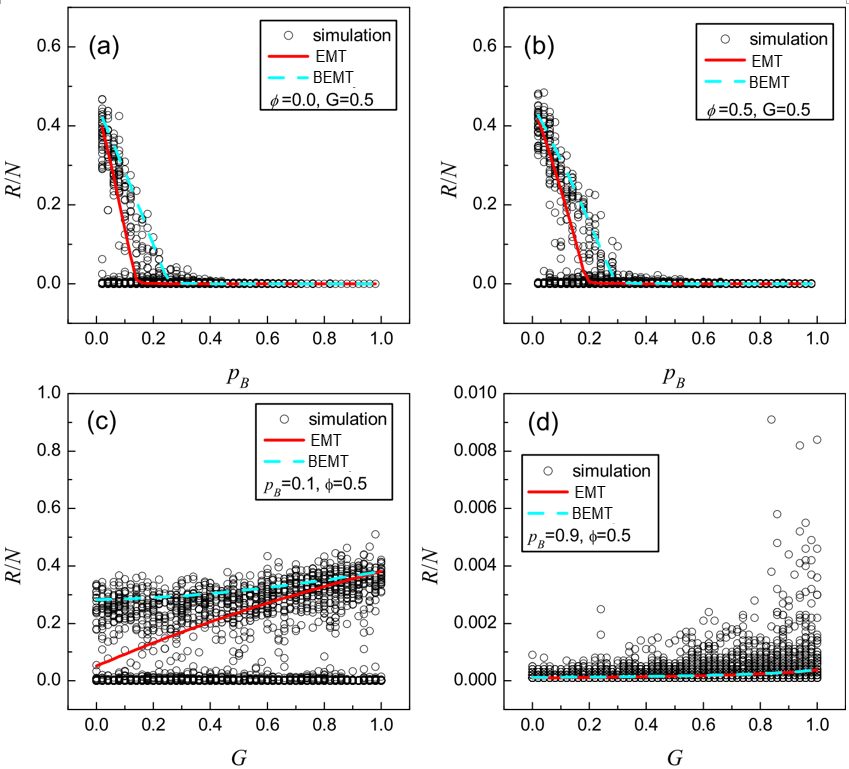,width=\linewidth} \caption{Simulation results for Case II with an immunity effect for the B species. The graphs describe the fraction of nodes reaching state R as a function of $p_B$ in (a) and (b) and as a function of $G$ in (c) and (d). The red solid curve and blue dash curve are for EMT and BEMT as defined in the theory section respectively. The parameters are: $N=10000$,
$\nu_c=0.2$, $\nu_f=0.03$, $\beta=0.01$, and $\gamma=0.05$.}
\label{fg:RvspBG}

\end{figure}

\subsection{Case III (Fig.~\ref{fg:RvspBF} and
Fig.~\ref{fg:RvspBG})}

To investigate the effects of an immunized subpopulation, we again choose 
an A node to be an I node at $t=0$. The
A node subpopulation is not immunized while the B node subpopulation 
is immunized. The immune effect is reflected by the parameter $\phi$ that relates $\beta$ and $\chi$ through $\chi = \beta \phi$.  Figure ~\ref{fg:RvspBF} shows some results when the cluster dynamics correspond to Case I.  For example, if the immune effect is
$100\%$ then there is no system-wide spreading if the
fraction of the immunized B population is above a critical value. For the cluster dynamics corresponding to Case II, some results are shown in Fig.~\ref{fg:RvspBG}. We specifically observed that as we increase the immune effect across the B species, the outcomes remained notably similar. While EMT tends to underestimate the results, BEMT tends to overestimate them -- probably because their respective averagings make then under-estimate and over-estimate correlations. It appears feasible to achieve no system-wide spreading at all for a given immune effect (see Fig.~\ref{fg:RvspBG} (d)).

\begin{figure}
\centering
\epsfig{figure=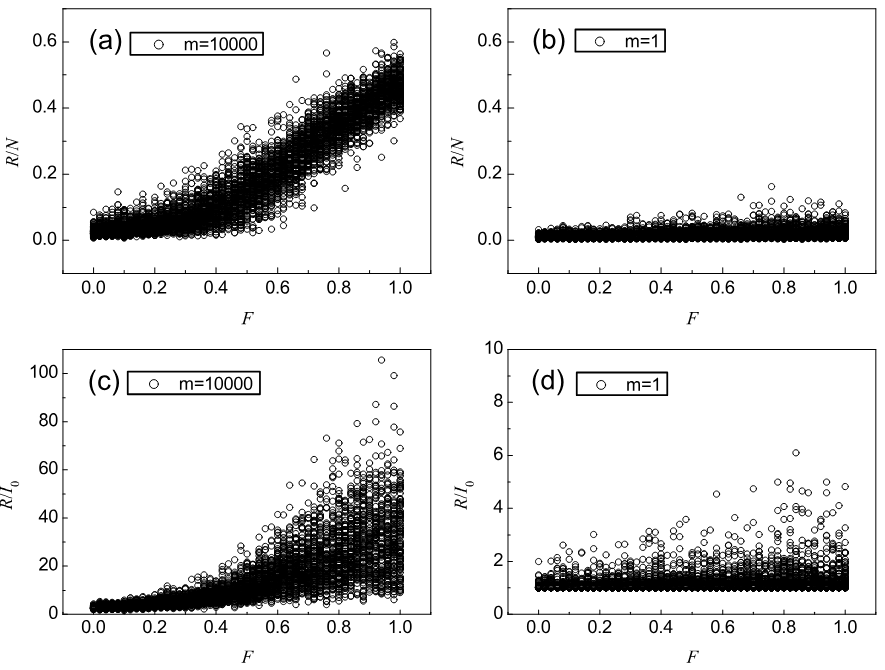,width=\linewidth} \caption{The fraction
of nodes reaching state R as a function of $F$ when $m=10000$ in (a) and (c) and $m=1$ in (b) and (d). The parameters are: $N=10000$,
$p_A=p_B=0.5$, $\nu_c=0.5$, $\nu_f=0.005$, $\beta=0.002$, and
$\gamma=0.015$.  The initial state has the biggest cluster all
infected.} \label{fg:allI}

\end{figure}

\begin{figure}[!t]
\centering
\epsfig{figure=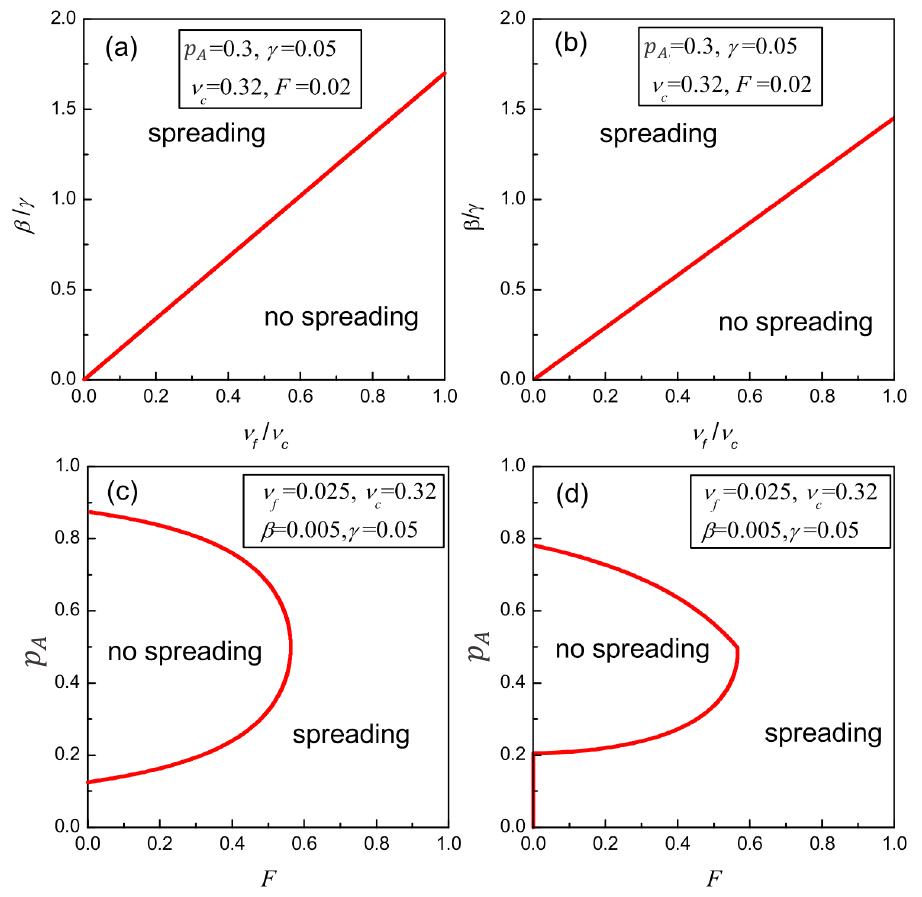,width=\linewidth}
\caption{Theoretical phase diagrams for Case I. (a) and (c) are
obtained by EMT.  (b) and (d) are obtained by BEMT. Other
parameters: $N=10000$. Initially, one node of species $A$ (hate node) is
infected.}  \label{fg:phsCaseI}

\end{figure}

\subsection{Spreading boundaries and phase diagrams}

The system-wide spreading--no-spreading phase boundary can be explored for the two theories, EMT and BEMT, by iterating the equations and seeing how $I(t)$ behaves over time. Then we check the results against simulation data. 
For the simulations, the stochastic nature of the SIR processes means that near the `boundary' between no-spreading and system-wide spreading, some simulation runs may yield system-wide spreading, and hence an epidemic outbreak, while others do not.  Even in an initial state of infecting all the nodes in the biggest cluster, this possibility of a stochastic evolution of the system to either a healthy state (no outbreak) or an epidemic outbreak state remains (see Fig.~\ref{fg:allI}). But visually, we can see that there is a threshold phenomenon. Hence, even though the continuous transition behavior plus the stochastic effect make it very difficult to determine the critical value precisely, the agreement with the theory curves is good. There is, therefore, some flexibility with the precise definition of the boundary between spreading and no-spreading, but the agreement with the theory is still good.

To obtain the phase diagram boundaries using the theories, we can define system-wide spreading (i.e., outbreak) and no-spreading (i.e., no outbreak) as follows. Let the initial number of infected nodes be $I(0)$. If the dynamical equations (see Sec. \ref{sec:theory}) give a result for $I(t)$ that exceeds $I(0)$, then we have a spreading state. If $I(t)\leq I(0)$ for all time with $I(t)\neq 0$, we have a no spreading state. Note that in iterating the equations, their continuous nature means that we could have $I(t)$ being any real number (not restricted to integers).  The idea is that the number of $I$ should keep on decreasing relative to $I(0)$ for a
no-spreading state. One may wonder about an unusual situation that could arise, whereby $I(0)$ is a large number and also there are large values of $\beta$ and $\gamma$ in the SIR process, hence even $I(t)<I(0)$ may result in the whole system being infected (thus becoming recovered) eventually. However this would be very rare -- and even in this case, the severity of infection is still decreasing when the population is infinitely large.

{\bf Case I.} Figure~\ref{fg:phsCaseI} shows the theoretical phase diagrams for Case I. The phase diagrams obtained by EMT
(Fig.~\ref{fg:phsCaseI}(a),(c)) and BEMT
(Fig.~\ref{fg:phsCaseI}(b),(d)) are qualitatively the same.  But
they show some quantitative differences that can be tested against
the simulation results. The slopes of the boundary in
Fig.~\ref{fg:phsCaseI}(a) and (b) are different.  EMT 
predicts a symmetric (parabolic-like) boundary
(Fig.~\ref{fg:phsCaseI}(c)) and BEMT  gives an asymmetric
shape.  If we fix $\nu_f/\nu_c=0.2$ and vary $\beta/\gamma$, we
will expect to see a transition from no spreading to spreading. If
we fix $\beta/\gamma=1.5$ and vary $\nu_f/\nu_c$, EMT 
(Fig.~\ref{fg:phsCaseI}(a)) predicts a spreading to no spreading
transition at $\nu_f/\nu_c$ around $0.9$, but BEMT  predicts a
spreading phase for the whole range of $\nu_f/\nu_c$.  If we fix
some value of $F$, e.g., $F=0.2$, and vary $p_{A}$, both EMT
and BEMT predict the system to go through a spreading to no-spreading and then  spreading transition, noting that the values
of $p_{A}$ at which the transitions occur are different. 

\begin{figure}[!t]
\centering
\epsfig{figure=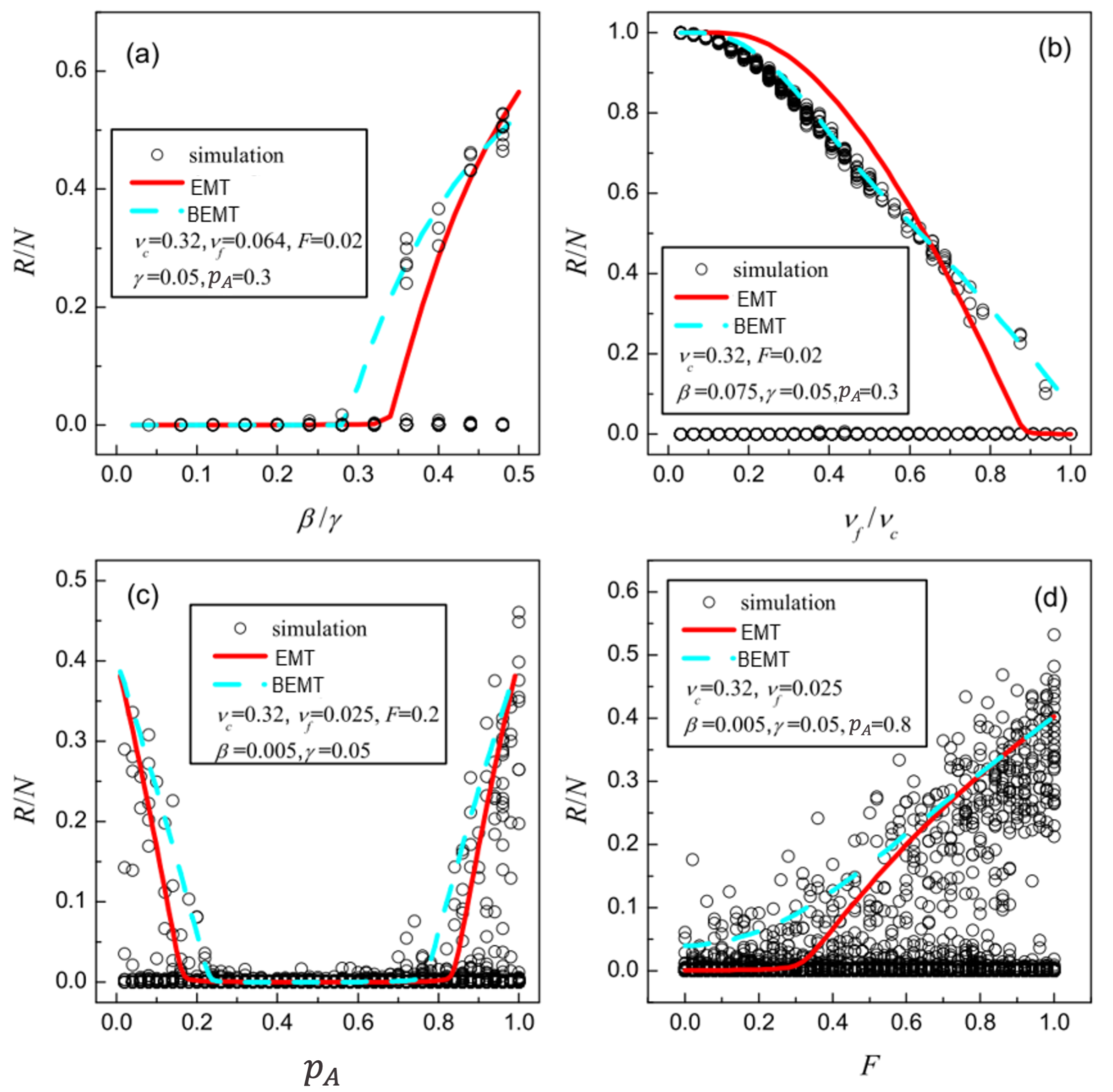,width=\linewidth}
\caption{Simulation results for the fraction of nodes reaching state R, choosing parameters based on the
phase diagrams for Case I. (a) corresponds to fixing $\nu_f/\nu_c
= 0.2$ (see Fig.~\ref{fg:phsCaseI}(a),(b)) and varying
$\beta/\gamma$.  (b) corresponds to fixing $\beta/\gamma = 1.5$ and varying $\nu_{f}/\nu_{c}$. (c) corresponds to fixing $F=0.2$ (see Fig.~\ref{fg:phsCaseI}(c),(d)) and varying $p_{A}$. (d) corresponds to fixing $p_{A} = 0.8$ and varying $F$.  Simulation data show spreading runs for all values of $F$.  As mentioned earlier, there are some no-spreading runs over 100 runs of the system. The red solid curve and blue dash curve are for EMT and BEMT as defined in the theory section respectively. Other parameters: $N=10000$ and $m=N$. Initially, one node of species A (hate node) is infected. The symbols are simulation data, with 100 runs for each set of variables (vertically, there are 100 symbols).}
\label{fg:CaseIsim-theo}

\end{figure}

Figure ~\ref{fg:CaseIsim-theo} shows the simulation results for Case
I with a comparison to the results of EMT and BEMT.  It should be read together with Fig.~\ref{fg:phsCaseI}.  In addition to the different phases, the results provide a quantitative comparison between EMT  and BEMT. Specifically, there is a no-spreading-to-spreading transition, as predicted by both theories in Fig. \ref{fg:CaseIsim-theo}a and Fig. \ref{fg:CaseIsim-theo}b.  BEMT gives better quantitative agreement for the $\beta/\gamma$ at which the transition occurs and the values of $R/N$ in the spreading phase. 

In Fig. \ref{fg:CaseIsim-theo}c, the simulation data show the re-entrant behavior of spreading $\rightarrow$ no spreading $\rightarrow$ spreading transitions. In Fig. \ref{fg:CaseIsim-theo}d,  EMT  predicts a transition from spreading to no spreading phase. However, BEMT predicts a spreading phase only. Simulation data show that BEMT works better in describing $R/N$ over the whole range of $\nu_{f}/\nu_{c}$. 

\begin{figure}
\centering
\epsfig{figure=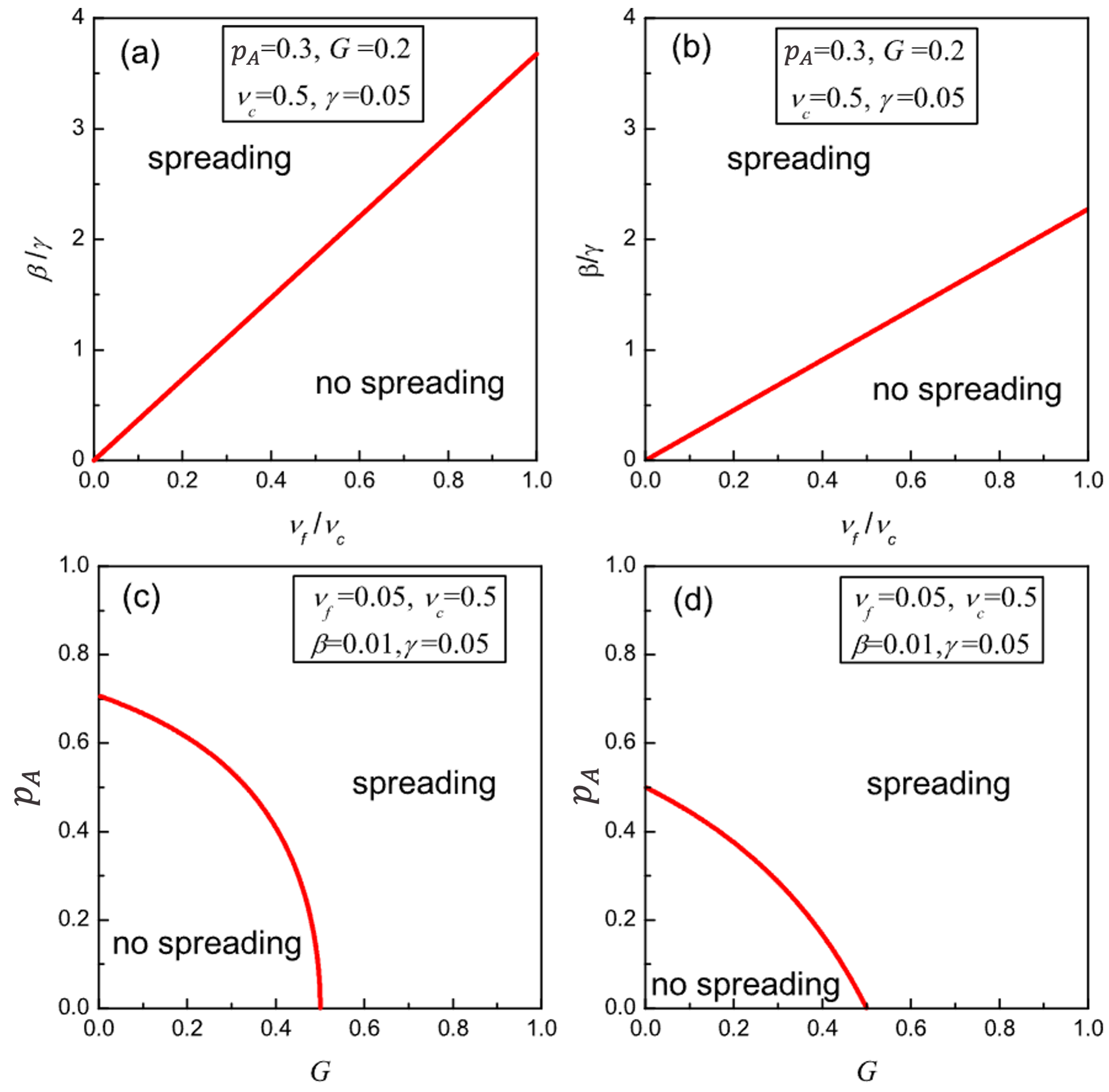,width=\linewidth}
\caption{Theoretical phase diagrams for Case II. (a) and (c) are
obtained by EMT. (b) and (d) are obtained by BEMT. Other
parameters: $N=10000$. Initially, one node of A species is
infected.}  \label{fg:phsCaseII}

\end{figure}

\begin{figure}
\centering
\epsfig{figure=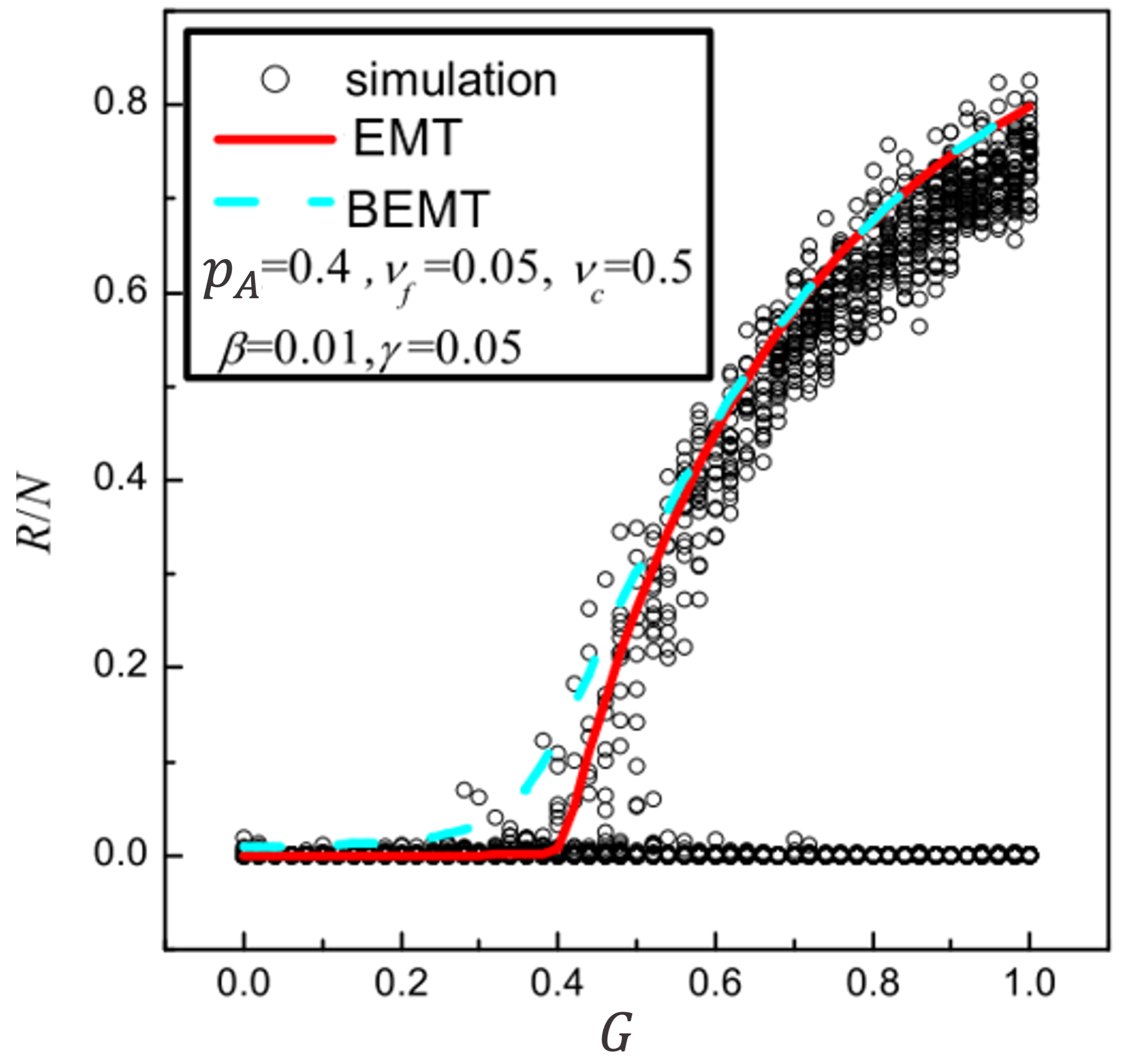,width=\linewidth}
\caption{Simulation results for the fraction of nodes reaching state R, choosing parameters based on the
phase diagrams for Case II.  The figure corresponds to fixing
$p_{A} = 0.4$ and varying $G$. The red solid curve and blue dash curve are for EMT and BEMT as defined in the theory section respectively. Other
parameters: $N=10000$ and $m=N$. Initially, one node of species A (hate node)
is infected. The symbols are simulation data, with 100 runs for
each value of $G$.} \label{fg:CaseIIsim-theo}

\end{figure}

\vskip0.1in
{\bf Case II.} Figure ~\ref{fg:phsCaseII} shows the theoretical phase boundary of
Case II obtained by EMT and BEMT.  The EMT and BEMT diagrams are qualitatively the same but have some quantitative differences along boundary conditions, similar to Case I.  Figure ~\ref{fg:CaseIIsim-theo} shows the
simulation results for fixing $p_{A} = 0.4$ and varying $G$. 

Unlike the boundary observed with $F$ (i.e. Case I) which shows the system transitioning from a spreading to non-spreading and to spreading pattern when the proportion of A species (hate nodes) is changed, we here typically observe a transition from a non-spreading to spreading pattern when the value of $G$ is fixed and the proportion of A species is increased. Both theories predict a no-spreading to spreading transition.  BEMT predicts a transition at a smaller value of $G$.  The data show that BEMT better captures $R/N$ in the range of $G \approx 0.3$ to $0.4$, where $R/N$ starts to increase.  For higher $G$,
the two theories agree with each other for $R/N$.
\vskip0.1in

\begin{figure}[!t]
\centering
\epsfig{figure=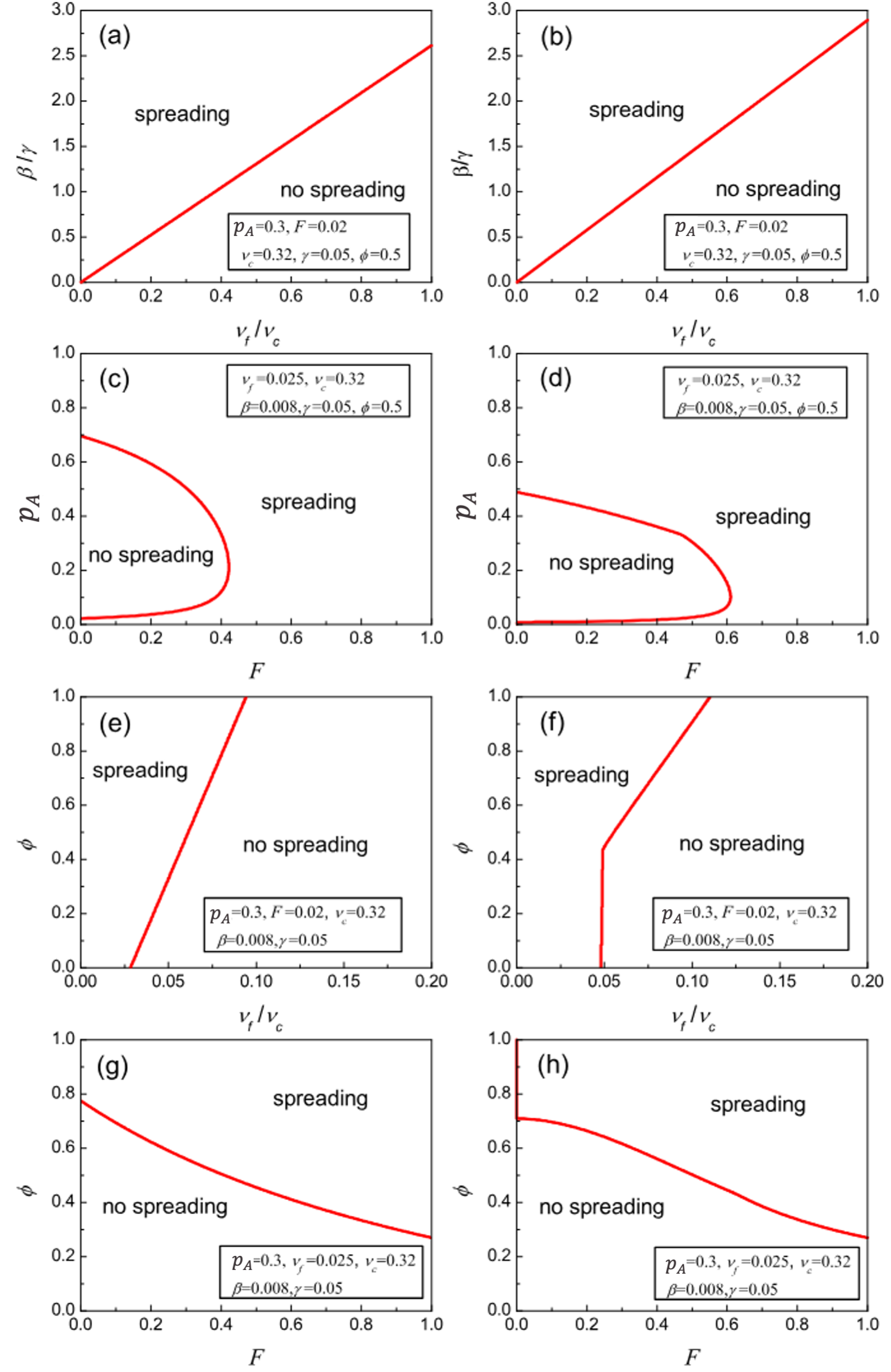,width=0.93\linewidth}
\caption{Theoretical phase diagrams for Case III when the contact
between A and B nodes belongs to Case I. (a), (c), (e), (g)
are obtained by EMT. (b), (d), (f), (h) are obtained by BEMT. Other parameters: $N=10000$. Initially, one node of species A (hate node)  is infected.} \label{fg:phsCaseIII-F}

\end{figure}

{\bf Case III.}  Figure~\ref{fg:phsCaseIII-F} shows the theoretical phase diagrams obtained by
EMT ((a),(c),(e),(g)) and BEMT ((b),(d),(f),(h)), when the contact between A and B nodes belongs to Case I, i.e., the two populations have same intra-coalesence probability rates and the inter-coalescence probability is a ratio of $F$ with them. 

\begin{figure}[!t]
\centering
\epsfig{figure=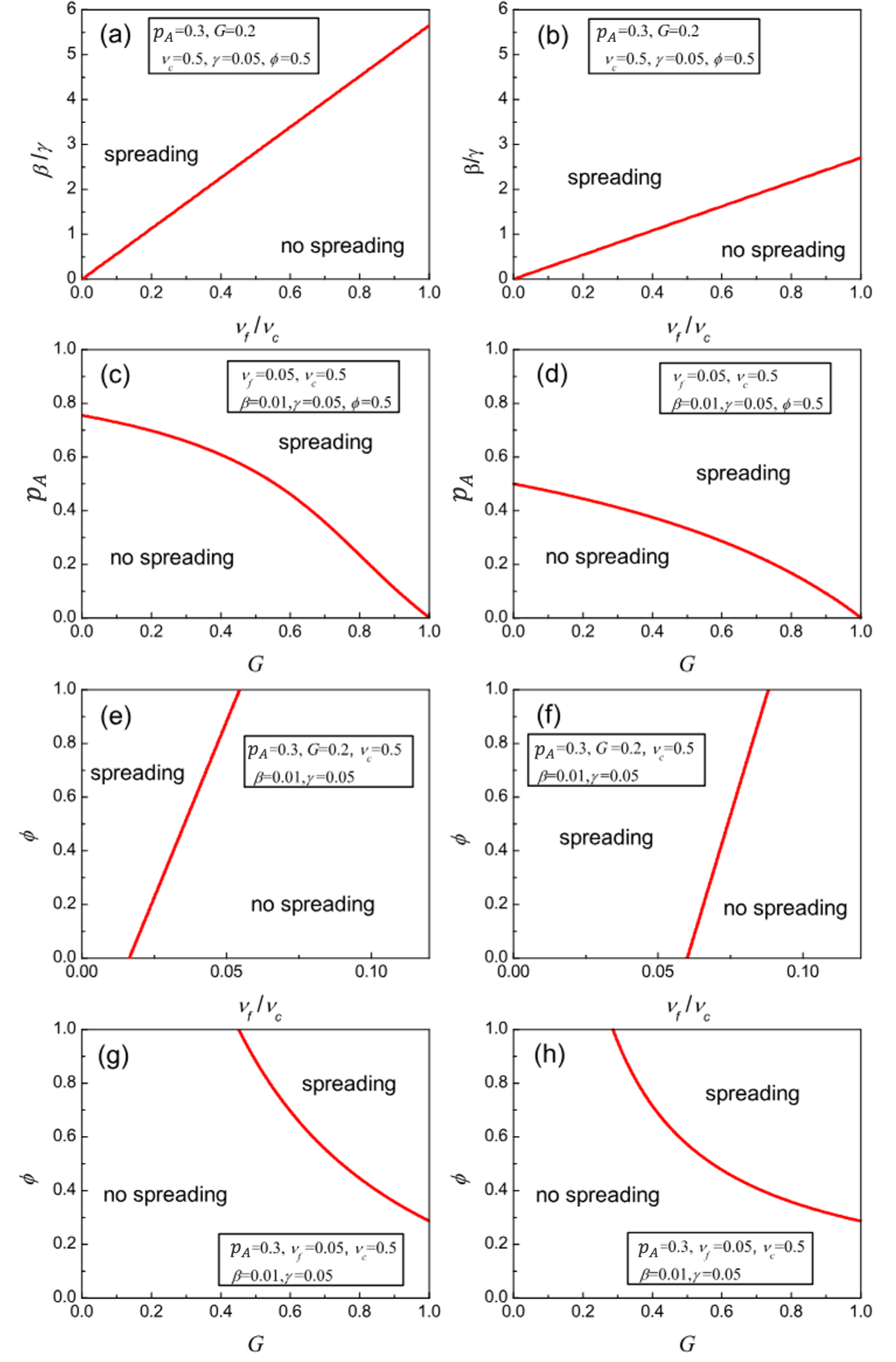,width=\linewidth}
\caption{Theoretical phase diagrams for Case III when the contact
between A and B nodes belongs to Case II. (a), (c), (e),
(g) are obtained by EMT. (b), (d), (f), (h) are obtained by
BEMT. Other parameters: $N=10000$. Initially, one node of
species A (hate node) is infected.}  \label{fg:phsCaseIII-G}

\end{figure}

\begin{figure}[!t]
\centering
\epsfig{figure=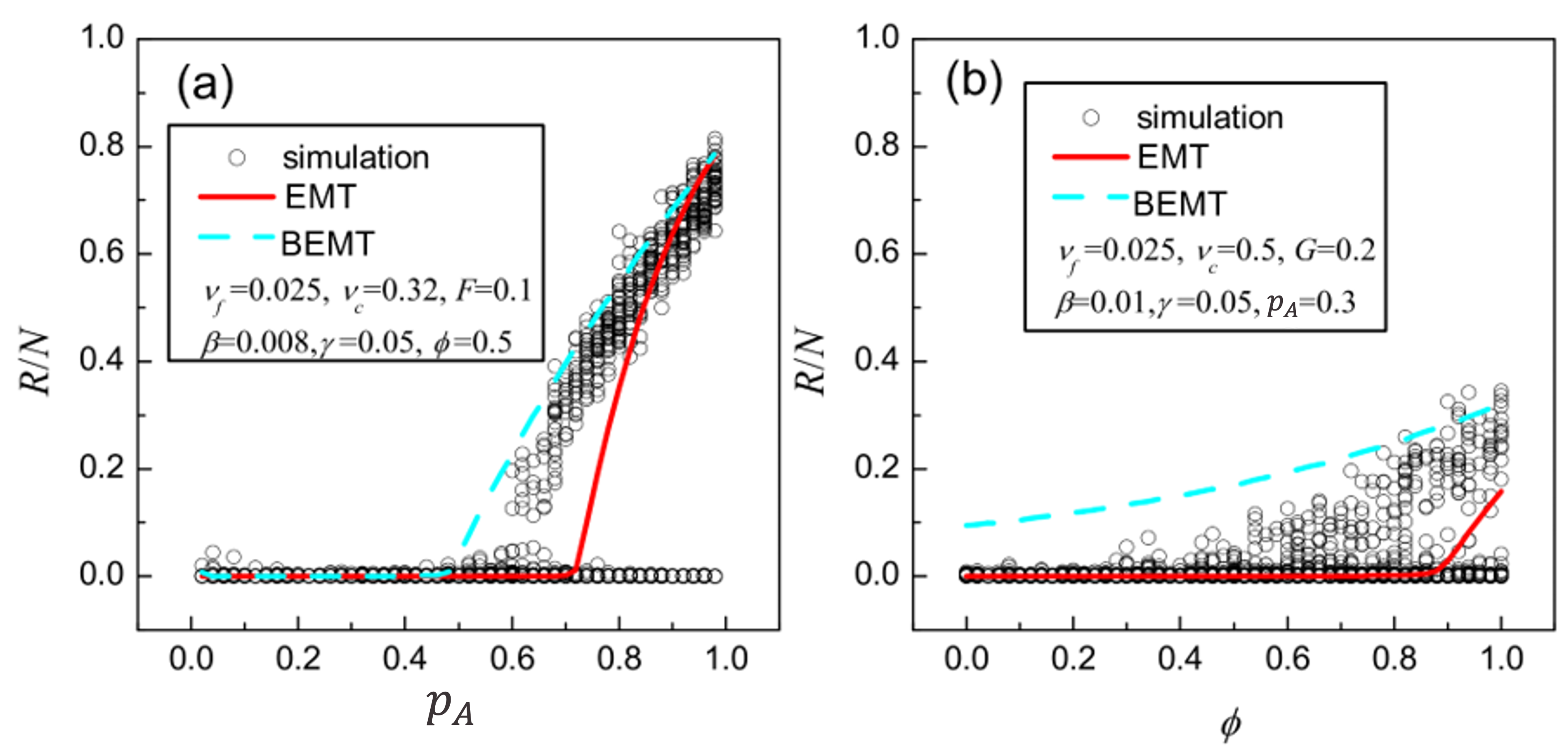,width=\linewidth}
\caption{Simulation results for the fraction of nodes reaching state R, choosing parameters based on the
phase diagrams for Case III. Other parameters: $N=10000$ and
$m=N$. Initially, one node of species A (hate node) is infected. The symbols
are obtained by simulations, with 100 runs for each set of
parameters.}  \label{fg:CaseIIIsim-theo}

\end{figure}

We observe very similar qualitative results between the theories.   Fig.~\ref{fg:phsCaseIII-G} shows the phase diagrams obtained by EMT ((a),(c),(e),(g)) and BEMT ((b),(d),(f),(h)), when the contact between A and B nodes belongs to Case II, i.e., the intra-coalescence probability are different between the two species with a ratio $G$. With one more parameter $\phi$, there are more features. Figure~\ref{fg:CaseIIIsim-theo} shows the simulation results for Case III.  In Fig.~\ref{fg:CaseIIIsim-theo}, the data are enclosed by the two theories: BEMT seems to overestimate $R/N$ and EMT underestimates $R/N$.

\begin{figure}
\centering
\epsfig{figure=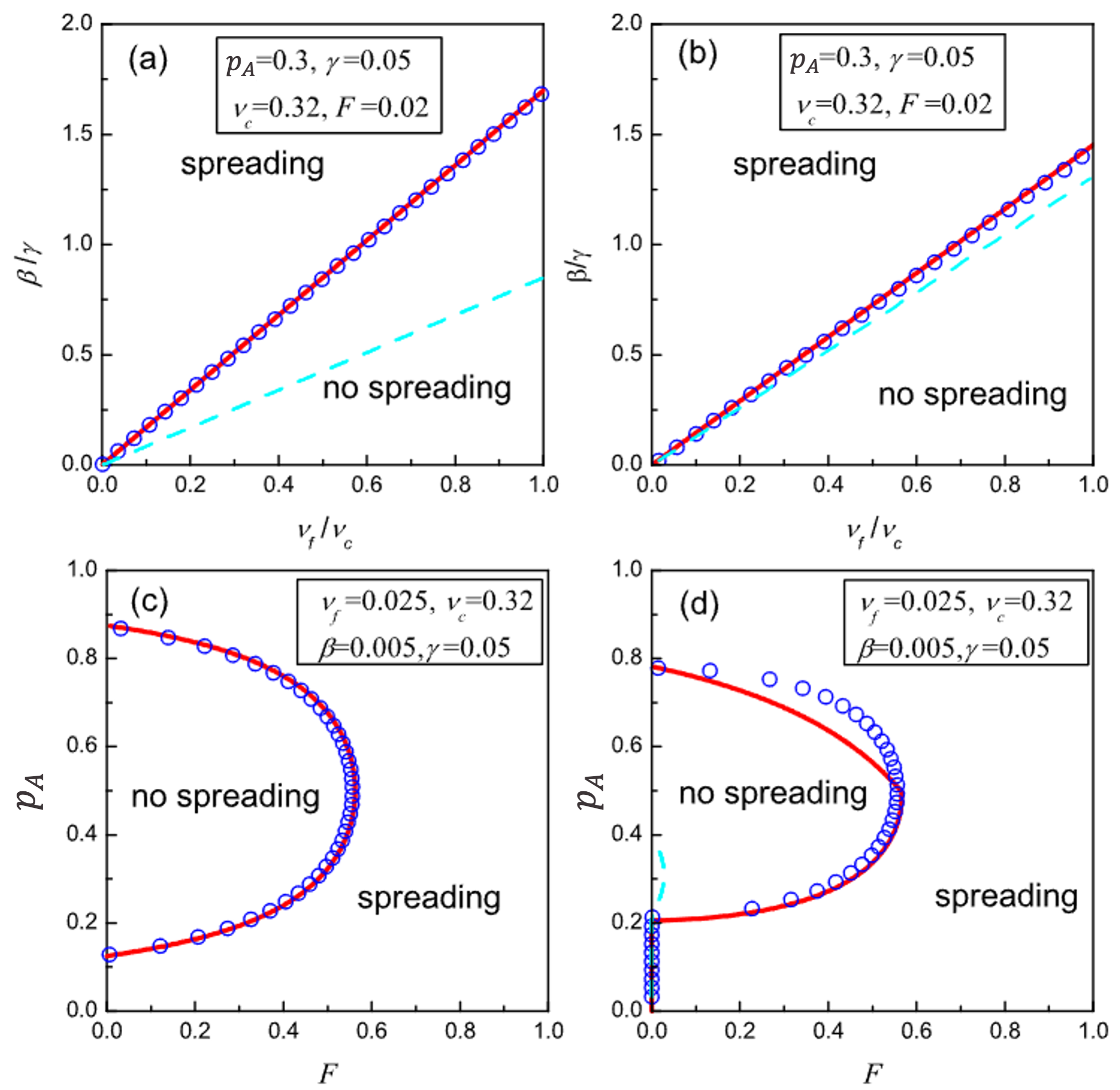,width=\linewidth}
\caption{Theoretical phase diagrams for Case I. (a) and (c) are
obtained by EMT. (b) and (d) are obtained by BEMT. The
solid curves (red online) are obtained by defining the spreading
threshold as $I_{max}>I(0)$ in the equations.  The circles are obtained by defining the spreading
threshold as $R(\infty) \geq 100$ (or $\Delta I  = 99$ and
$I(0)=1$) in the simulations. Other parameters: $N=10000$. Initially, one node of
species A (hate node) is infected. As a further comparison, the dashed (cyan) curves show the spreading threshold if one were to use the criterion $R(\infty)\geq
I(0)+1$.} \label{fg:phsCaseIB}

\end{figure}

\bigskip
Finally, Fig.~\ref{fg:phsCaseIB} compares the phase boundaries from theory and simulation for Case I. We note that EMT and BEMT become equivalent when $p_{A} = p_{B} = 1/2$, $F=1$, and $N p_{A} \gg 1$.


\section{Policy Implications and Conclusion}

Having presented all the details of the theoretical and numerical calculations in previous sections, we will dedicate this section to drawing together the key findings to generate broader takeaways that could impact policy discussions. We will make this section stand-alone in the sense that it does not require a detailed knowledge of prior sections, and instead provides a fairly self-contained summary of how our findings are relevant to policymakers.

\begin{figure}[!t]
\epsfig{figure=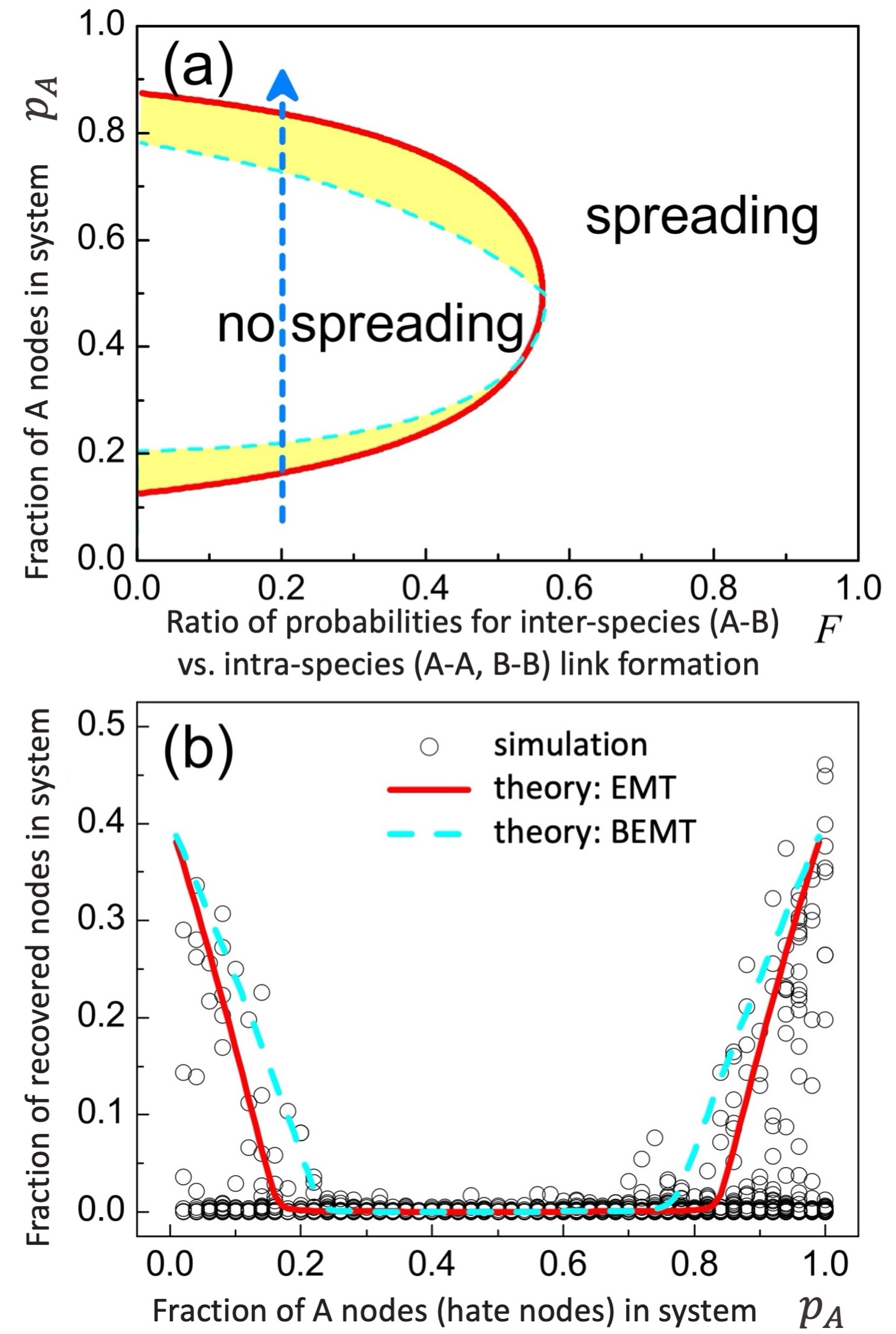,width=0.85 \linewidth}
\caption{Impact on system-wide spreading when the probability of inter-species link formation is a fraction $F<1$ of the probability of intra-species link formation (i.e. Case I). (a) Phase diagram for system-wide spreading. In Figs. \ref{fg:Figure2}-\ref{fg:Figure4}, the red (blue) curve is EMT (BEMT) theory which tends to underestimate (overestimate) correlations. Hence the yellow area is a crude indicator of spread in simulation output. $\nu_f=0.025$, $\nu_c=0.32$, $\beta=0.005$, $\gamma=0.05$. In Figs. \ref{fg:Figure2}-\ref{fg:Figure4}, $N=10000$, $m=N$, and one A node (hate node) is infected at $t=0$ with link dynamics already in steady state. (b) Simulation and theory for $F=0.2$ and increasing $p_A$, which corresponds to dotted vertical line in (a).}  \label{fg:Figure2}
\end{figure}
\vskip0.1in

First, we review our equations' predictions for the empirically relevant scenario where the probability of {\em inter}-species link appearances differ from {\em intra}-species link appearances (Case I). Specifically, $\nu_{cAB}=\nu_{cBA} = F \nu_c$ with 
$\nu_{cAA}=\nu_{cBB}=\nu_c$, and $\nu_{fA}=\nu_{fB}=\nu_f$. $0\leq F < 1$ means the probability of link formation is smaller for nodes from different species than for the same species.
Figure \ref{fg:Figure2} shows the theoretical phase boundaries and illustrates the good agreement with the numerical simulation of the model. The simpler EMT appears sufficient for this specific parameter range, though Sec.  \ref{sec:simulation} confirms that the BEMT is better elsewhere in the parameter space. 
As expected, both theories predict that one can prevent spreading from species A  by reducing $F$ -- but non-trivially, this is {\em only} true if both A and B have comparable fractions of nodes ($p_A$ not too far from $0.5$). Specifically, if (and only if) $p_A$ falls within the semi-ellipse shown, species A can have a finite $F$, yet system-wide spreading will still not occur.

\begin{figure}
\centering
\epsfig{figure=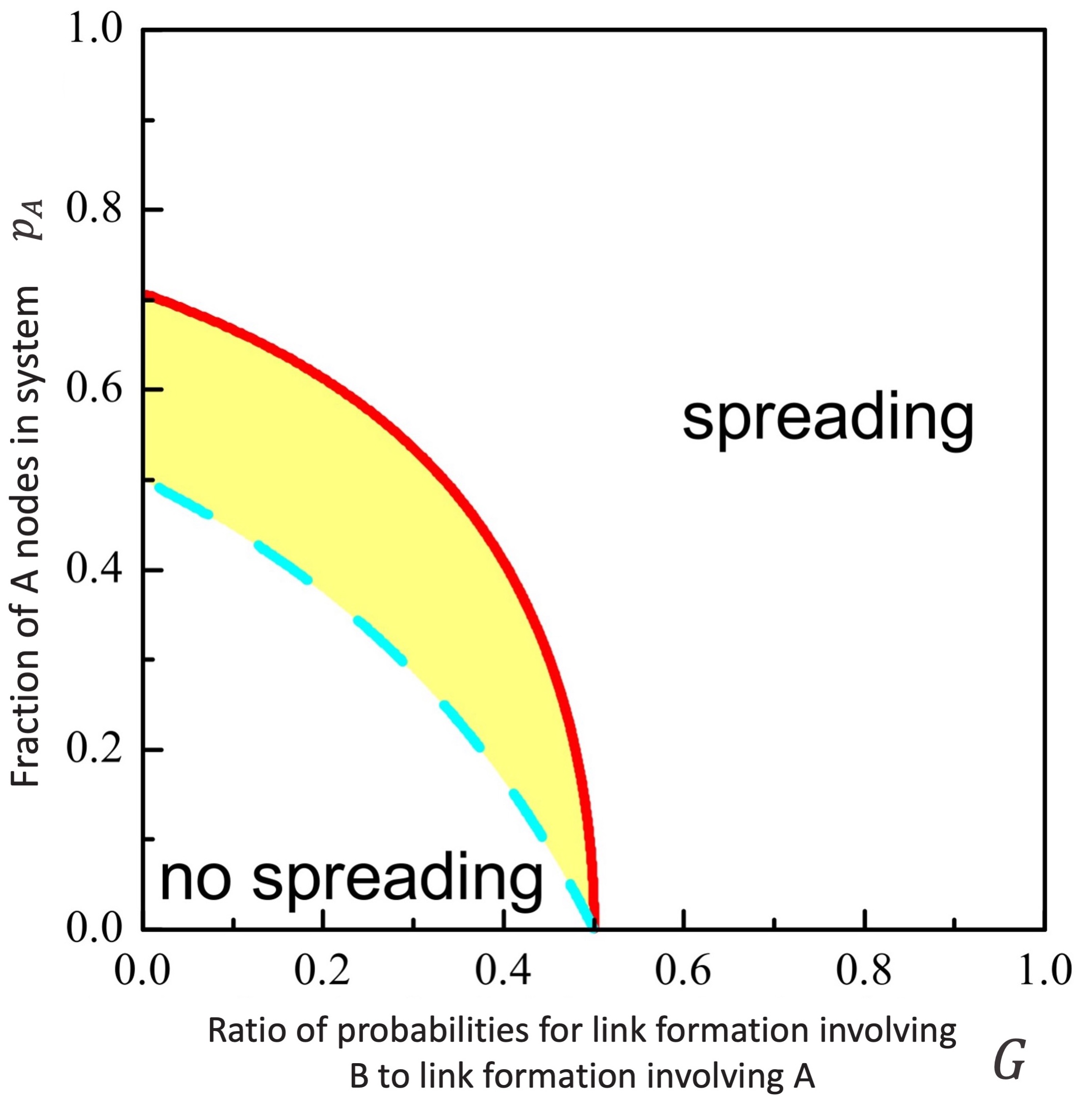,width=0.65\linewidth}
\caption{Impact on system-wide spreading when the probability of link formation involving species B (i.e. non-hate nodes) is a fraction $G<1$ of the probability of link formation within species A (i.e. hate nodes) only, i.e. Case II. Phase diagram for system-wide spreading.  $\nu_f=0.05$, $\nu_c=0.5$, $\beta=0.01$, $\gamma=0.05$. 
}  \label{fg:Figure3}

\end{figure}
\begin{figure}
\centering
\epsfig{figure=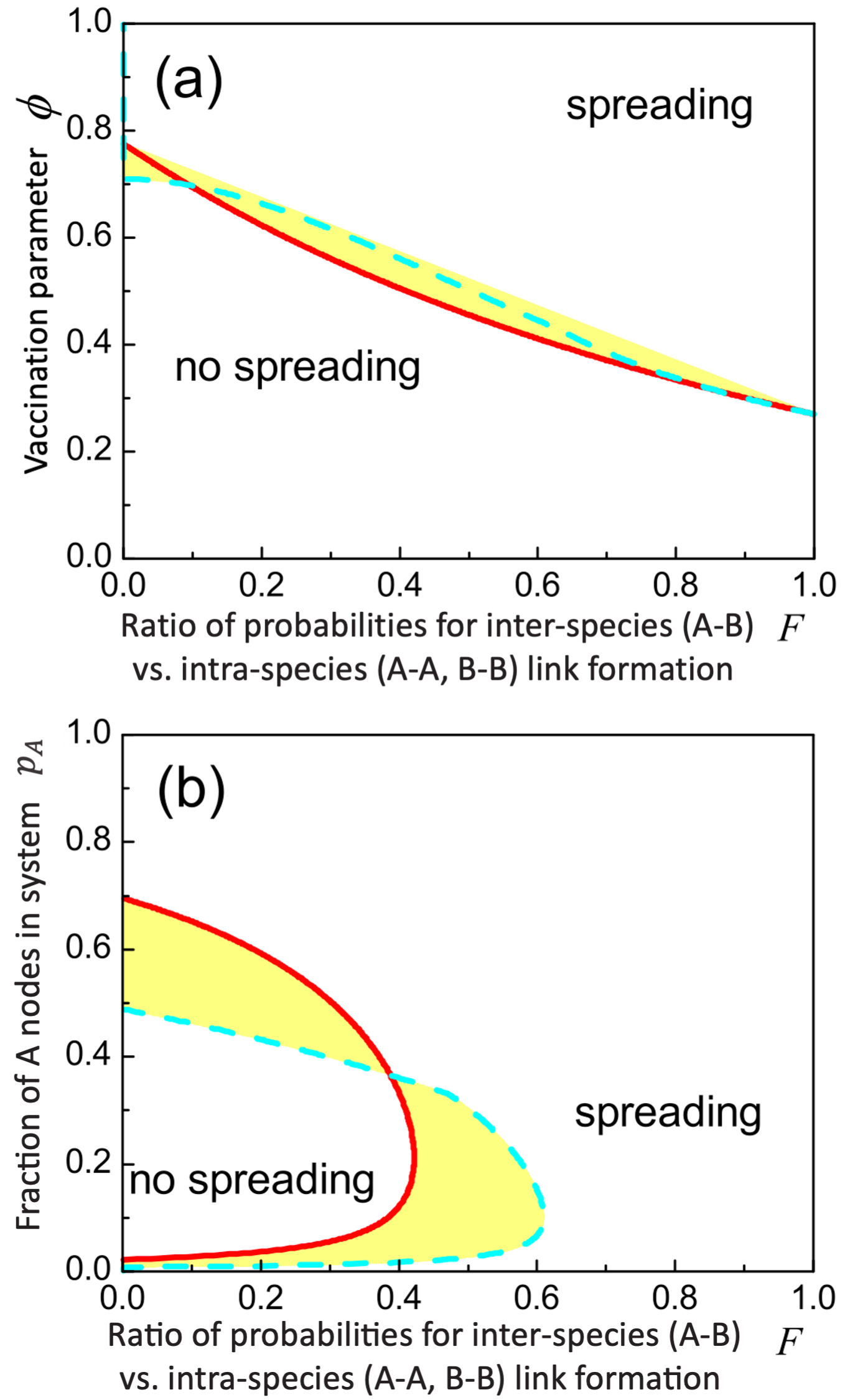,width=0.85\linewidth}
\caption{Impact of online vaccination on system-wide spreading. Here, the vaccinated B nodes infection rate is now $\chi=\phi\beta<\beta$. (a) The boundary shows the $\phi$ values required to achieve no system-wide spreading for a given $F$, i.e., to achieve online herd immunity. Link probabilities as in Case I, with $p_A=0.3$. (b) Vaccination distorts phase diagram compared to Fig.~\ref{fg:Figure2}(a). $\phi=0.5$, $\nu_f=0.025$, $\nu_c=0.32$, $\beta=0.008$, $\gamma=0.05$.} \label{fg:Figure4}
\end{figure}

Figure \ref{fg:Figure2}(a) also suggests that with $F$ fixed and small, and $p_A$ evolving so slowly that the theories remain valid (i.e., clustering-viral dynamics much faster than changes in $p_A$), the system-wide spreading will exhibit re-entrant phases (vertical blue line). 

\vskip0.1in
{\em Physical explanation of re-entrant behavior.} The re-entrant phase behavior arises from the competition between two effects as $p_A$ changes. When $p_A$ is very small, few A nodes are embedded in large B-dominated clusters through which content can spread system-wide. When $p_A$ is very large, A nodes dominate and form large clusters enabling spreading. At intermediate $p_A$, neither species alone sustains sufficiently large clusters because the effective coalescence rate $\tilde{\nu}_c$ (Eq.~2) depends non-linearly on $p_A$ via the weighted sum of intra- and inter-species coalescence rates. The SIR reproduction number $\kappa$ (Eqs.~\ref{eq:kappa1}, \ref{eq:kappa02}) therefore has a non-monotonic dependence on $p_A$, creating the re-entrant gap. Importantly, the BEMT predicts an {\em asymmetric} phase boundary even for Case~I (Fig.~\ref{fg:phsCaseI}(d)), because the initial infection is placed in an A node, breaking the A$\leftrightarrow$B symmetry. Case~II is explicitly asymmetric and still exhibits non-trivial threshold structure (Fig.~\ref{fg:phsCaseII}), confirming that the phenomenon is not a trivial symmetry artifact.

\vskip0.1in
This re-entrant behavior warns that if the hate species A moves into a no-spreading regime as $p_A$ increases from a small initial value, it can suddenly tip to a spreading regime as $p_A$ gets large. Likewise if A has a reduction in its number of nodes (i.e., $p_A$ decreases from a large initial value) and enters a no-spreading phase, it can suddenly tip to a spreading phase as $p_A$ gets further reduced. 

\vskip0.1in
{\em This re-entrant behavior hence yields a simple example where platforms doing `more' of the same makes things worse}. For example, species A may have been steadily experiencing a reduction in the number of its communities -- hence $p_A$  was reducing from an initial large value. The outcome for society looked good initially when it went from a spreading phase to a no-spreading phase, i.e. traversing the vertical arrow in reverse in Fig.~\ref{fg:Figure2}(a) from top to bottom. But then when $p_A$ reduced further, the system suddenly jumped back to a spreading phase. This outcome resulting from the re-entrant behavior shown in Fig.~\ref{fg:Figure2}(a), shows that ordering social media platforms to do more of the same to control content spreading (i.e. keep banning problematic communities and hence keep reducing $p_A$), can end up suddenly making the outcome worse.

Second, we considered our equations' predictions for the empirically relevant scenario of species-dependent frequency of link appearances (Case II).
We mimicked this by setting $\nu_{cAB}=\nu_{cBA}=\nu_{cBB}=G\nu_c$ with
$\nu_{cAA}=\nu_c$. $0\leq G \leq 1$ means B nodes are less likely to form links as compared to A nodes. Figure \ref{fg:Figure3} shows the theoretical phase boundaries: 
no-spreading requires small $G$ and small $p_A$, which makes a dominant (i.e. large $p_A$) A species {\em less likely to be able to prevent spreading} irrespective of how it controls its frequency of link appearances relative to other species.

Third, our analysis quantified the online herd immunity required to prevent system-wide spreading (Case III). The concept of a `digital vaccination' has already been pursued by some social media companies including Meta (Facebook): it posts positive counter-messages on Facebook pages (nodes) to try to prevent misinformation from taking hold. Vaccinated nodes  have infection probability $\beta$ reduced to $\chi=\phi\beta$ where $0\leq\phi<1$. So ${\dot { S}_A(t)}  = -\beta P {S}_A(t) {I}(t)$, ${\dot {S}_V(t)}  = -\chi P {S}_V(t) {I}(t)$,  ${\dot {I}}(t)  = P (\beta {S}_A(t)  + \chi {S}_V(t) ){I}(t) - \gamma {I}(t)$ within EMT ($V\equiv B$ hence $p_\mathrm{n}\equiv p_V$ etc.).
Hence the critical fraction of all $N$ nodes that need to be vaccinated to prevent system-wide spreading is:
\begin{equation}
\label{eq:equation6}
p_V \geq \f{1}{1-\phi} \L 1-\f{\gamma}{PN\beta}\R \ . 
\end{equation}
\noindent Equation \ref{eq:equation6} also applies when it is only the B nodes that are all vaccinated, i.e., $p_B\equiv p_V$ is fixed, in which case Eq. \ref{eq:equation6} predicts the level of vaccine efficacy $(1-\phi)$ required to prevent system-wide spreading. Figure~\ref{fg:Figure4}(a) shows the boundary given by solving for $\phi$ in the equality in Eq. \ref{eq:equation6}. The shift to higher $F$ as $\phi$ decreases means that increasing vaccine efficacy $(1-\phi)$ allows higher probabilities of inter-species link formation $F$ to be tolerated while still avoiding system-wide spreading. Figure \ref{fg:Figure4}(b) is similar to Fig. \ref{fg:Figure2}(a), but shows the impact of having all B nodes vaccinated, i.e. the phase boundaries are distorted, but the size of the spreading region does not change much. 

This provides another warning: the inter-species link dynamics play such a crucial role that vaccination of one species or platform can simply end up refocusing the harmful material toward another. 

\vskip0.2in
To conclude, we have presented a two-species coalescence-fragmentation model with SIR dynamics for the spreading of hate content online. As discussed in Secs. I and II, this study fills a gap left open by three prior works: the single-species coalescence-fragmentation--SIR model of Ref.~\cite{usPRE2010} (which was purely numerical with no analytic theory or systematic phase diagram exploration), the coalescence-fragmentation framework of Ref.~\cite{PRL2023} (no SIR), and the $D$-species generalization of Ref.~\cite{PRL2024} (no SIR). The present paper combines multi-species ($D=2$) coalescence-fragmentation dynamics with SIR contagion, derives analytic mean-field theories at two levels of approximation (EMT and BEMT), and systematically maps the resulting phase diagrams -- discovering re-entrant spreading phases that were absent from all three prior works. The derived analytic formulae give explicit insight into how the phase boundaries might be manipulated to prevent system-wide spreading. 
More broadly, the re-entrant spreading phases provide a specific, qualitatively robust prediction: that steadily deplatforming hate communities can initially succeed but then backfire beyond a critical threshold. We stress that this qualitative prediction holds across broad parameter ranges, though the precise quantitative thresholds depend on specific parameter values that would need to be determined empirically for any given scenario.

Given the simplicity of our model analysis, much remains to be done. The `infection' terminology is imperfect, and more complex viral processes may apply. But our mathematical results can be easily generalized to multiple species. We think that the immunity aspect is of particular promise, and we point to Ref. \cite{Om} in which the author worked with one of us to lay out an initial theoretical framework that later led to this in-depth study. 

Overall, our results show how physics thinking can play an important role in the mitigation of hate content spreading -- at the very least, by providing a concrete starting point for framing discussions and, hence, moving beyond vaguer verbal debates.

\begin{acknowledgments}
We are grateful to Nicholas J. Restrepo, Rhys Leahy, Nicolas Velasquez and Yonatan Lupu for help in obtaining the empirical data in Fig. 1, and King Yan Fong for additional help. NFJ is grateful to Om Kant Jha for their joint work together during his PhD thesis which opened up a pathway to this more general in-depth study.
\end{acknowledgments}

\section*{Data Availability Statement}
The data in Fig. 1 have already been published in Ref. \cite{Multi2021}. Data output from the simulations and information about the simulation software, is available from the authors on reasonable request. An interactive simulation dashboard for the coalescence-fragmentation dynamics is freely accessible at \url{https://gwdonlab.github.io/netlogo-simulator/} \cite{netlogo} from which the source code can be downloaded.

\appendix

\section{Expressions for link probabilities in BEMT}
\label{app:A}

We can break $P$ up exactly into

\begin{equation}
\label{eq:Pexact}
  P = P_{AA}  + P_{AB}  + P_{BB}  
\end{equation}

where the quantities of $P_{AA}$, $P_{AB}(\equiv P_{BA})$ and $P_{BB}$ are
\begin{equation}
P_{AA}  = \frac{{\nu_{cAA} p_A^2 }}{{\nu_{cAA} p_A^2  +
(\nu_{cAB}  + \nu_{cBA} )p_B p_A  + \nu_{cBB} p_B^2 }}P
\;,\label{eq:meanPHH}
\end{equation}

\begin{equation}
P_{AB}  = \frac{{(\nu_{cAB}  + \nu_{cBA} )p_B p_A
}}{{\nu_{cAA} p_A^2  + (\nu_{cAB}  + \nu_{cBA} )p_B p_A  +
\nu_{cBB} p_B^2 }}P \;,\label{eq:meanPHN}
\end{equation}

\begin{equation}
P_{BB}  = \frac{{\nu_{cBB} p_B^2 }}{{\nu_{cAA} p_A^2  +
(\nu_{cAB}  + \nu_{cBA} )p_B p_A  + \nu_{cBB} p_B^2 }}P \;.
\label{eq:meanPNN}
\end{equation}

\vspace*{0.3 true in}

As in the main paper, we now consider two different cases of 2-species 
parameter regimes, to characterize the impact of the inhomogeneity (A and B).

\noindent {\bf Case I:} $\nu_{fA}=\nu_{fB}=\nu_f$,
$\nu_{cAA}=\nu_{cBB}=\nu_c$, $\nu_{cAB}=\nu_{cBA} = F \nu_c$. The
parameter $F$ sets the across-species cluster coalescence with
respect to the single-species cluster coalescence.  For Case I, Eq.
(\ref{eq:Pexact}) and Eqs. (\ref{eq:meanPHH})-(\ref{eq:meanPNN}) become

\begin{eqnarray}
  P &=& \frac{{\nu _c \left( {p_A^2  + 2Fp_B p_A  + p_B^2 } \right)}}{{\nu _c \left( {p_A^2  + 2Fp_B p_A  + p_B^2 } \right) + N\nu _f }}, \label{eq:meanPI} \\
P_{AA} &=& \frac{{p_A^2 }}{{p_A^2  + 2Fp_B p_A  + p_B^2 }}P,\label{eq:meanPHHI} \\
P_{AB} &=& \frac{{2Fp_B p_A }}{{p_A^2  + 2Fp_B p_A  + p_B^2 }}P,\label{eq:meanPHNI}\\
P_{BB} &=& \frac{{p_B^2 }}{{p_A^2  + 2Fp_B p_A  + p_B^2 }}P. \label{eq:meanPNNI}
\end{eqnarray}

\noindent {\bf Case II:} $\nu_{fA}=\nu_{fB}=\nu_f$,
$\nu_{cAA}=\nu_c$, $\nu_{cAB}=\nu_{cBA}=\nu_{cBB}=G\nu_c$.  Here,
one species (B) has a different coalescence probability with
respect to the other species.  For Case II, Eq. (\ref{eq:Pexact}) and
Eqs. (\ref{eq:meanPHH})-(\ref{eq:meanPNN}) become
\begin{eqnarray}
 P &=& \frac{{\nu _c [ {p_A^2  + (2p_A p_B  + p_B^2 )G} ]}}{{\nu _c [ {p_A^2  + (2p_A p_B  + p_B^2 )G} ] + N\nu _f }}, \label{eq:meanPII}\\
P_{AA} &=& \frac{{p_A^2 }}{{p_A^2  + (2p_A p_B + p_B^2 )G}}P,\label{eq:meanPHHII} \\
P_{AB} &=& \frac{{2p_B p_A G}}{{p_A^2  + (2 p_A p_B + p_B^2 )G}}P,\label{eq:meanPHNII}\\
P_{BB} &=& \frac{{p_B^2 G}}{{p_A^2  + (2 p_A p_B + p_B^2 )G}}P. \label{eq:meanPNNII}
\end{eqnarray}

\vspace*{0.2 true in}

Comparing the values obtained from the numerical simulations to these expressions, we find that the crudest level approximation for $P$ works well
in both cases. However, the more granular forms $P_{AA}$, $P_{AB}$, $P_{BB}$ work well in some ranges of the parameters but not all.
For
Case I and Case II when the parameters $F$ and $G$ are close to 1, 
these are the cases where the inhomogeneity in
the system is small, i.e. B type and A type coalescences are not too
different.  In these cases, the mean field expressions
$P_{AA}$, $P_{AB}$ and $P_{BB}$ are very close to the numerical results.
This situation is like the one in standard binary alloy problems.  When the two atoms behave very differently, then a simple average (as a linear function connecting the $F=0$ ($G=0$) and $F=1$ ($G=0$) limits) is not very accurate, as in the so-called virtual
crystal approximation: i.e. when two types of atoms differ significantly
(either in terms of on-site energies or hopping integrals), then some more complex approximations would be needed.

\section{Alternative derivation of $P$ starting from clustering equations}

Here we give a more microscopic derivation of the expression for $P$. For simplicity of notation, we limit ourselves to a single species type or averaging regime where there are species-independent coalescence and fragmentation probabilities $\nu_{c}$ and $\nu_{f}$. We show below that by applying averaging to the clustering equations directly -- which is a different averaging approach to that adopted in the main paper -- the same expression emerges for $P$ in the limiting case $N\nu_{f}\gg\nu_{c}$. This hence illustrates the robustness of our expression for $P$. 

We start by considering the master equation for the number $n_s$ of clusters-of-nodes with size $s$ in the model:
\begin{equation}\label{ezmaster1}
\frac{\partial n_{s}}{\partial
t}= - \frac{\nu_{f}sn_{s}}{N} + \frac{\nu_{c}}{N^{2}}\sum_{s'=1}^{s-1}s'n_{s'}(s-s')n_{s-s'}
-  \frac{2\nu_{c}sn_{s}}{N^{2}}\sum_{s'=1}^{\infty}s'n_{s'}\end{equation}
for $s\geq 2$, with a similar but truncated form for $s=1$:
\begin{equation}\label{ezmaster2}
\frac{\partial n_{1}}{\partial
t}=\frac{\nu_{f}}{N}\sum_{s'=2}^{\infty}(s')^{2}n_{s'}-\frac{2\nu_{c}n_{1}}{N^{2}}\sum_{s'=1}^{\infty}s'n_{s'}\ \ .
\end{equation}
For a steady-state distribution, we have
\begin{equation}\label{ezstable}
sn_{s}=\frac{\nu_{c}}{2\nu_{c}+\nu_{f}}\frac{1}{N}\sum_{s'=1}^{s-1}s'n_{s'}(s-s')n_{s-s'}
\end{equation}
for $s\geq2$, while for $s=1$ we have
\begin{equation}\label{ezstable1}
n_{1}=\frac{\nu_{f}}{2\nu_{c}}\sum_{s'=2}^{\infty}(s')^{2}n_{s'}\ \ .
\end{equation}
Therefore on average, we obtain:
\begin{equation}\label{prob}
P=\sum_{s=2}^{\infty}\frac{sn_{s}}{N}\frac{s-1}{N}=\frac{1}{N^{2}}\sum_{s=2}^{\infty}(s^{2}n_{s}-sn_{s})
=\frac{1}{N^{2}}\frac{2\nu_{c}}{\nu_{f}}n_{1}-\frac{N-n_{1}}{N^{2}}
\end{equation}
where the only unknown quantity is $n_{1}$. We now define a generating function:
\begin{equation}\label{eqgen}
G[y]=\sum_{k=0}^{\infty}kn_ky^k=n_1y+\sum_{k=2}^{\infty}kn_ky^k=n_1y+g[y]\ \ .
\end{equation}
Multiplying by $y^{s}$ and then summing
from $s=2$ to $\infty$, yields
\begin{equation}
g[y]=\frac{\nu_{c}}{2\nu_{c}+\nu_{f}}\frac{1}{N}G[y]^2
\end{equation}
i.e.,
\begin{eqnarray}
g[y]^{2}-(\frac{2\nu_{c}+\nu_{f}}{\nu_{c}}N-2n_{1}y)g[y]+n_{1}^{2}y^{2}=0
\end{eqnarray}
where $g[y]=\sum_{s=2}^{\infty}sn_{s}y^{s}$ and $g[1]=N-n_{1}$.
Solving this quadratic equation gives
\begin{eqnarray}
n_{1}=\frac{\nu_{f}+\nu_{c}}{\nu_{f}+2\nu_{c}}N\ .
\end{eqnarray}

Substituting into Eq. \ref{prob}, we obtain the expression for $P$ in the limit $N\nu_{f}\gg\nu_{c}$, i.e. $P=\nu_c/N \nu_f$. For the more general case of 
species-dependent coalescence and fragmentation probabilities
as in the main paper, a similar derivation follows and this equation becomes 
$P = {\tilde \nu _c}/N\tilde \nu _f$
 which is exactly the same as Eq. 7 in the main paper.

\begin{figure*}[!t]
\centering
\includegraphics[width=0.75\textwidth]{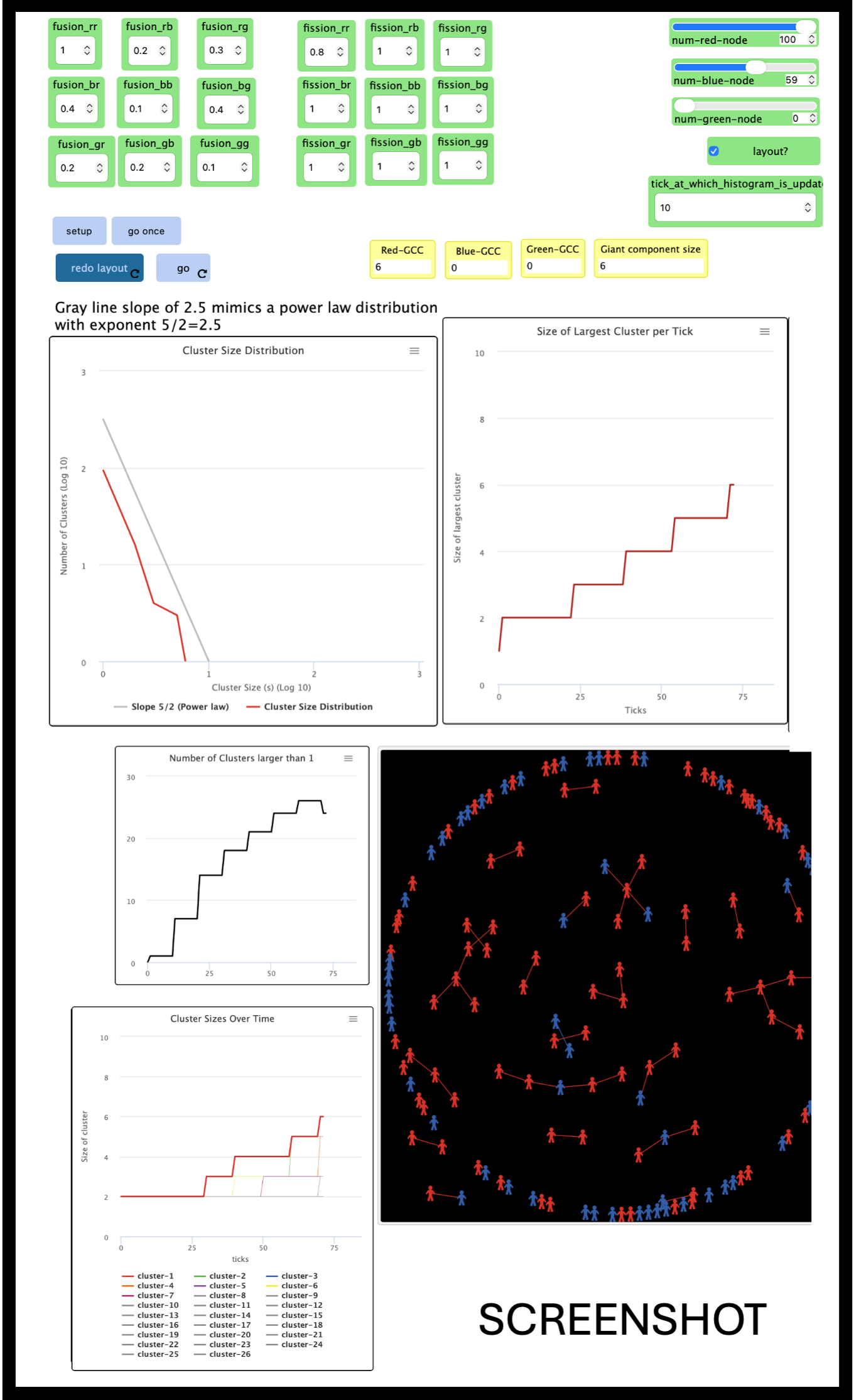}
\caption{Screenshot from our free, publicly accessible interactive simulation dashboard (\url{https://gwdonlab.github.io/netlogo-simulator/} from which the source code can be downloaded) showing the coalescence-fragmentation dynamics for two species (A and B, with third species population set to zero). Upper-right: real-time network visualization showing dynamically evolving clusters of mixed species composition. Lower-left: log-log cluster size distribution with $-5/2$ power-law reference line (gray). Lower-center: size of the largest cluster vs.\ time, showing the characteristic fluctuations of the dynamical steady state. Lower-right: individual cluster sizes tracked over time, showing coalescence (size jumps up) and fragmentation (size drops to zero) events. All coalescence (fusion) and fragmentation (fission) probabilities are independently adjustable via the control panel (upper left).}
\label{fg:netlogo}
\end{figure*}

\end{document}